\documentclass[aos, preprint]{imsart}

\RequirePackage{amsthm,amsmath,amsfonts,amssymb}
\RequirePackage[authoryear]{natbib}
\usepackage{epsfig,graphics}
\usepackage{amssymb, amsmath,array}
\usepackage{natbib}
\usepackage{bm,url}
\usepackage{color}
\usepackage{calligra}
\usepackage{longtable,booktabs,array}
\usepackage{booktabs}
\usepackage[normalem]{ulem}

\RequirePackage[colorlinks,citecolor=blue,urlcolor=blue]{hyperref}
\usepackage[capitalize]{cleveref}

\usepackage{xcolor}
\theoremstyle{plain}

\newtheorem{theorem}{Theorem}[section]

\newtheorem{prop}{Proposition}
\newtheorem{exam}{Example}

\theoremstyle{definition}

\def \E {{\rm E}}

\def \P {\mathbb{P}}
\def \R {\mathbb{R}}
\def \H {\mathbb{H}}

\def \rI {{\rm I}}
\def \bco{\iffalse}

\usepackage{amsmath, amsthm, amssymb, amsfonts}
\DeclareMathOperator*{\argmin}{arg\,min}

\newcommand{\hf}{{\hat f}}
\newcommand{\hs}{{\hat s}}

\newcommand{\hM}{{\hat M}}

\newcommand{\tf}{{\tilde f}}

\newcommand{\hcT}{{\hat{\cal T}}}
\newcommand{\tcT}{{\tilde{\cal T}}}

\newcommand{\hatm}{{\hat m}}

\newcommand{\hpi}{{\hat \pi}}
\newcommand{\htau}{{\hat \tau}}
\newcommand{\ttau}{{\tilde \tau}}

\newcommand{\tpi}{{\tilde \pi}}

\newcommand{\tm}{{\tilde m}}
\newcommand{\tM}{{\tilde M}}

\newcommand{\bX}{{\bf X}}
\newcommand{\bx}{{\bf x}}

\newcommand{\bu}{{\bf u}}

\newcommand{\bb}{{\bf b}}

\newcommand{\bz}{{\bf z}}

\newcommand{\cH}{{\cal H}}

\newcommand{\cM}{{\cal M}}

\newcommand{\cT}{{\cal T}}
\newcommand{\cS}{{\cal S}}

\newcommand{\F}{Fr\'echet }
\newcommand{\Om}{\Omega }
\newcommand{\om}{\omega }
\newcommand{\OL}{\Omega_{\mathcal{L}}}

\newcommand{\hw}{{\hat w}}

\def\ci{\cite}
\def\cp{\citep}
\def\bs{\boldsymbol}

\newcommand{\dx}{\,{\mathrm d}\bx}

\newcommand{\dxj}{\,{\mathrm d}x_j}
\newcommand{\dxk}{\,{\mathrm d}x_k}

\startlocaldefs

\endlocaldefs

\begin{document}

\begin{frontmatter}
\title{Additive \F Regression for Random Objects}
\runtitle{Additive \F Regression}

\begin{aug}
\author[A]{\fnms{Changwon}~\snm{Choi}\thanks{Equal contribution}\ead[label=e1]{cwchoi94@gmail.com}},
\author[B]{\fnms{Wookyeong}~\snm{Song}\protect\footnotemark[1]\ead[label=e2]{wksong@ucdavis.edu}}
\author[B]{\fnms{Hans-Georg}~\snm{M\"uller}\ead[label=e3]{hgmueller@ucdavis.edu}}
\and
\author[A]{\fnms{Byeong U.}~\snm{Park}\ead[label=e4]{bupark@snu.ac.kr}}

\address[A]{Department of Statistics, Seoul National University\printead[presep={ ,\ }]{e1,e4}}

\address[B]{Department of Statistics,
University of California, Davis\printead[presep={,\ }]{e2,e3}}
\end{aug}

\begin{abstract}
Regression analysis for complex data taking values in a general metric space  has gained increasing attention in recent years, particularly in the context of \F regression with Euclidean predictors $\bX \in \R^{p}$. However, local \F regression, while more flexible than global \F regression,  suffers from the curse of dimensionality for the case of multivariate predictors. To address this issue while maintaining model flexibility, we introduce a framework of additive structured nonparametric regression  when responses are situated in a general metric space. Overcoming the challenge of the
lack of vector space structure in general metric spaces where the responses reside requires  a novel approach  that  incorporates the additive structure via projection operators. This leads to  a unified framework of additive regression for a wide range of response types, including one-dimensional distributions in Wasserstein space, network data represented by graph Laplacians and spherical data equipped with geodesic distances, among others.  We establish pointwise and uniform consistency, including  convergence rates, for the proposed additive \F regression estimators, using  smooth backfitting. The practical utility of this additive model is demonstrated through applications to brain connectivity network analysis using resting-state fMRI data and  distributional physical activity data.
\end{abstract}


\end{frontmatter}


\section{Introduction}

In metric statistics and object data analysis one considers statistical methodology for data referred to as random objects that reside in a general metric space $(\Omega, d).$  This area is receiving  increasing attention due to the growing prevalence of complex data  that can be viewed as elements of a metric space. Such random objects include samples of networks that are encountered in neuroimaging to model brain connectivity \citep{fornito2016fundamentals, zhou2022network}, probability distributions equipped with the Wasserstein geometry \citep{mull:23:2}, compositional data \citep{zhu2023autoregressive} and symmetric positive definite  matrices equipped with various metrics  \citep{lin2023additive}.

A basic task in this context is modeling the regression relationship between a random object response $Y$ in $(\Omega, d)$ and a  Euclidean predictor  $\bX$ in $\R^{p}$. If $p>1,$ i.e., the predictor  $\bX$ is multivariate,  more flexible approaches such as local \F regression \citep{chen2022uniform} are subject to the curse of dimensionality of $\bX$ and therefore dimension reduction is  required. The goal is to maintain as much flexibility as possible while avoiding the curse of dimensionality. To achieve this with an explicitly specified model,  for the classical case of Euclidean responses there are  two established major approaches: Single index regression \citep{lin2007identifiability} and additive regression \citep{buja1989linear}.
For the single-index model and related dimension reduction approaches, extensions to accommodate random object responses have recently been developed \citep{zhang2024dimension,bhattacharjee2025nonlinear,bhattacharjee2023single,ghosal2023frechet, hong2025robust}.

For additive modeling with metric space-valued responses,  we are only aware of an extrinsic approach in \cite{yang2025variable}. This approach was developed independently and concurrently with our work and has an
objective that is quite different from ours, namely to identify informative covariates. Importantly, the variable selection is in terms of estimating the conditional mean of the difference between the squared distances of $Y$ from two fixed points in $\Omega$, and thus the focus differs substantially from ours, as we aim to  develop  a comprehensive intrinsic  framework for  additive regression for random object responses. As far as we know, such a framework has remained unexplored to date. In this paper we
take up the challenge of developing such an intrinsic extension of additive modeling in the absence of linear operations in the metric space where the responses reside.

The  main innovation of this paper is that for the first time a model with intrinsic additive structure in the response space is developed and studied in the absence of linear operations, where no intrinsic additive model is currently available. Notably, the proposed approach  provides an intrinsic approach, while previous approaches for additive modeling for random object responses targeted specific types of random objects, for example  distributional responses, and relied on transformations to a linear space such as a Hilbert space and therefore were inherently extrinsic \citep{han2020additive}.
These prior approaches depend on specifically tailored transformations that map data into a suitable linear space, where linear operations can be deployed to implement an additive model in the transformed space. In contrast, our method does not require any such transformation. A key limitation of  transformation-based  extrinsic methods is that they rely on a non-invertible embedding map into a linear Hilbert space. It is therefore not guaranteed that the fitted responses can be represented in the original metric space and therefore their interpretation may be
imperiled \citep{mull:23:2,pegoraro2022projected}.

Recent studies have explored additive regression for random object responses in a few specific metric spaces, including Hilbert space-valued responses \citep{jeon2020additive}, distribution-valued responses \citep{han2020additive}  and symmetric positive-definite matrix-valued  responses \citep{lin2023additive}. We develop here   a novel unifying framework for intrinsic additive regression models when responses are situated in a general metric space. The key innovations of this paper can be summarized as follows: \begin{itemize}
    \item We  provide a general intrinsic framework for additive regression that can be applied to general metric space-valued responses, including  networks and compositional data.
    \item We extend smooth backfitting to object responses when fitting   additive Fr\'echet regression models.
    \item We  provide a comprehensive asymptotic analysis including consistency and convergence rates for the estimators of the components of the additive model  for  general metric response spaces.
    \item We demonstrate the practical utility of the proposed additive \F regression model for relevant applications. For the case of distributional responses, the proposed  intrinsic additive model is shown to improve upon the  extrinsic additive model in \cite{han2020additive}.
\end{itemize}

Additive  \F regression targets   conditional \F means when the effect of the predictors is restricted to have an  additive structure in an implicit manner. For Euclidean responses,  smooth backfitting (SBF),   a kernel-based method for estimating additive structures that was  originally developed by \cite{mammen1999existence},  has proved to be a powerful approach for additive modeling. It has been recently extended to Hilbert space-valued responses \citep{jeon2020additive}. Although the SBF framework has been adapted to various structured nonparametric regression models \citep{cho2024partially,yu2008smooth,lee2012flexible,bissantz2016smooth}, its application to object-valued responses in general metric spaces has remained unexplored so far; here we address this gap.

Our first data illustration involves the analysis of brain connectivity networks from individuals diagnosed with Alzheimer’s disease and cognitively normal controls, using multivariate Euclidean explanatory variables such as cognitive assessment scores and protein biomarker levels. The second application focuses on distributional physical activity data collected with  accelerometers, along with demographic, dietary and clinical covariates. These examples show the flexibility of the new additive  framework for modeling object-valued responses while addressing the challenges posed by multivariate  predictors.

The remainder of this paper is organized as follows. In Section \ref{sec:add_Freg_model}, we introduce the additive \F regression model, along with motivating examples of random object responses in metric spaces.
Estimation procedures and implementation details for the SBF approach are provided in Sections \ref{sec:Add-F-reg-mean-est} and \ref{sec:net_nu_imp}.
Asymptotic properties,  including consistency and convergence rates for the additive \F regression estimators obtained by using the SBF algorithm,  are introduced in Section \ref{sec:asymptotics}. Simulation studies involving network and univariate density responses are presented in Section \ref{sec:simul}. Section \ref{sec:data_application} illustrates the proposed methodology using resting-state fMRI data from the ADNI study and distributional activity data from the  NHANES database. We conclude with a brief discussion in Section \ref{sec:discussion}. Auxiliary results and  proofs are provided in the Supplementary Material \citep{choi2026AFR}.

\section{Additive \F regression}\label{sec:add_Freg_model}

\subsection{Preliminaries: Additive models for Euclidean responses}

Let $\bX$ be a $p$-dimensional covariate vector taking values in $[0,1]^p$ and $Y$  a response variable taking values in a metric space $(\Omega,d)$.
Throughout this paper, we assume that the number of covariates, $p$, is fixed.
We study a regression model with $\bX$ as  covariate vector and $Y$ as response.
Denote by $f$, $f_j$ and $f_{j,k}$ the density functions
of $\bX \equiv (X_1, \ldots, X_p)^\top$, $X_j$ and  $(X_j,X_k)$ with respect to the Lebesgue measures on $\R^p$, $\R$ and $\R^2$, respectively.
The space of additive regression functions when $\Omega=\R$ is
\begin{align}\label{H-add-sp}\begin{split}
\cH_+^\R&:=\{g \in L_2^\R(f)\,|\,g(\bx)=g_1(x_1)+\cdots+g_p(x_p) \;\mbox{for univariate} \\
&\mbox{functions}\; g_j:[0,1]\to \R\}.
\end{split}
\end{align}
Here and throughout, $L_2^\R(q)$ for a given density $q$ on $[0,1]^p$ is the space of functions $g:[0,1]^p \to \R$ such that $\int_{[0,1]^p}g(\bx)^2 q(\bx)\,\dx<\infty$  and
$\cH_j^\R:=\{g \in L_2^\R(f)\,|\,g(\bx)=g_j(x_j)\;\mbox{for some}\; g_j:[0,1]\to \R\}$ are  subspaces of $\cH_+^\R$, where
$\cH_+^\R=\cH_1^\R+\cdots+\cH_p^\R$.

In the following, we often interpret univariate functions $g_j$, which are the constituents  of an additive function $g_+: [0,1]^p \to \R$,
as multivariate functions $\tilde g_j: [0,1]^p \to \R$ such that
$\tilde g_j(\bx)= g_j(x_j)$  and  do not distinguish $\tilde g_j$ from $g_j$ in our notation.
With this convention, we  write $g_+=g_1+\cdots+g_p$. For $g \in L_2^\R(q)$, we write $g=0$ if $\int g(\bx)^2q(\bx)\dx=0$.
Also, $g=\eta$ for $g, \eta \in L_2^\R(q)$ means $g-\eta=0$.

When $\Omega=\R$, consider the projection of $m_\R:=\E(Y|\bX=\cdot)$ onto $\cH_+^\R$, which we denote by $m_{+,\R}$.
Let $m_{j,\R} \in \cH_j^\R$ for $1 \le j \le p$ be a set of component functions such that
$m_{+,\R}=m_{1,\R}+\cdots +m_{p,\R}$.
Since $m_\R-m_{+,\R} \perp \cH_j^\R$ for all $1 \le j \le p$,
\[
\E(Y-m_{+,\R}(\bX)|X_j=x_j)=0, \quad 1 \le j \le p.
\]
From this we see that any set of components $(m_{j,\R}:1\le j \le p)$ that compose $m_{+,\R}$ satisfies
\begin{align}\label{sbf-eq-pop}
m_{j,\R}(x_j)=\E(Y|X_j=x_j)-\sum_{k=1,\neq j}^p \int_0^1 m_{k,\R}(x_k)\frac{f_{j,k}(x_j,x_k)}{f_j(x_j)}\dxk, \quad 1 \le j \le p.
\end{align}

\subsection{Additive structure for \F regression}\label{sec:Add-F-reg-mean}

To prepare for the extension of  the definition of the additive function $m_{+,\R}$ to a general metric space $(\Omega, d)$,  where
 a vector space structure is not available, we require an alternative formulation of $m_{+,\R}$,  the solution
of \eqref{sbf-eq-pop}.  This alternative approach is motivated by the \F regression principle and related approaches \cp{petersen2019frechet,schotz2022nonparametric}.
The starting point for these approaches  is  an established estimator for real-valued responses that is linear in the data. Given data $(\bX_i,Y_i)$, where $\bX_i \in \R^p, \, Y_i \in \R$, such estimators are of the form
$\hat{\E}(Y|\bX=\bx)= \sum_{i=1}^n w(\bx,\bX_i) Y_i$ for  suitable weight functions $w:\, \R^p \times  \R^p \rightarrow \R$ satisfying $\sum_{i=1}^n w(\bx,\bX_i)=1$ and include local and global linear regression,
kernel estimators, B splines, smoothing splines and other commonly used estimators for the regression $\E(Y|\bX=\cdot)$.
As an  example, in the special case of  basic Nadaraya-Watson kernel estimators one would have
$w(\bx,\bX_i)=\frac{\prod_{j=1}^pK((x_j-X_{ij})/h)}{\sum_{i} \prod_{j=1}^pK((x_j-X_{ij})/h)}$, where the kernel $K$ is a probability density function and $h>0$ a bandwidth.
In general, the weights $w(\bx,\bX_i)$ need not be positive and indeed do include negative weights for the case of local and global linear regression.
Once a weight function $w$ has been determined, the extension to the case where $Y \in (\Om,d)$ is then obtained by
$\hat{\E}(Y|\bX=\bx)= \text{argmin}_{\om \in  \Om}  \sum_{i=1}^n w(\bx,\bX_i) d^2(\om, Y_i).$
We note that  this is a strict extension of the classical estimators, since they re-emerge   for the special case when $\Om=\R$ with the Euclidean metric  $d_E(y,a)=|y-a|$. Just
like their Euclidean counterparts, the local unrestricted  versions of \F regression are subject to the curse of dimensionality, which limits their applicability when there are  two or more
predictors.

In the following, we will apply the above \F  extension principle to additive regression modeling, which then requires specifying a weight function that implements the additive structure,
and we aim to develop a smooth backfitting approach to fit the model.
A first step is to develop the weight function perspective for the case of additive regression for Euclidean responses with the
Euclidean metric, which in itself provides a novel perspective.

For a baseline kernel function $K$, let $K_h: [0,1]\times [0,1] \to \R$ be a normalized weight function defined by $K_h(z,u)=c_h(u) h^{-1}K((u-z)/h)$,
where $c_h(u)$ is a normalizing constant to ensure $\int_0^1 K_h(z,u)\,{\rm d} z=1$ for all $u \in [0,1]$.
Consider an approximation $\tm_{+,\R}$ of $m_{+,\R}$  as
a solution of \eqref{sbf-eq-pop} when $f_j, \, f_{j,k}$ and $\E(Y|X_j=\cdot)$ are replaced by their `smoothed' versions, $\tf_j:=\E K_h(\cdot,X_j)$,
$\tf_{j,k}:=\E K_h(\cdot,X_j)K_h(\cdot,X_k)$ and $\E\{s_j(\cdot,\bX,h)Y\}$, where
\[
s_j(\bx,\bu,h)\equiv s_j(x_j,\bu,h)\equiv s_j(x_j,u_j,h):=\frac{K_h(x_j,u_j)}{\E K_h(x_j,X_j)}.
\]
The system of equations for $\tm_{+,\R}=\tm_{1,\R}+\cdots +\tm_{p,\R}$ is then
 determined by $s_j,\, \tf_j$ and $\tf_{j,k}$ as follows:
 \begin{align}\label{tilde}
\tm_{j,\R}(x_j)= \E\{ s_j(x_j,\bX,h)Y\}  -\sum_{k=1,\neq j}^p \int_0^1 \tm_{k,\R}(x_k)\frac{\tf_{j,k}(x_j,x_k)}{\tf_j(x_j)}\dxk, \quad 1 \le j \le p.
\end{align}

Define
\[
\tf(\bx):=\E (\prod_{j=1}^p K_h(x_j,X_j)),
\]
where in  $\tf_j, \, \tf_{j,k}$ and $\tf$ we  suppress the dependence on the bandwidth $h$.
For each $\bx\in[0,1]^p$, the map $s_j(\bx,\cdot,h):[0,1]^p\to\R$ is an element of $L_2({\rm unif})$,
where $L_2({\rm unif})$ denotes the space of square integrable real-valued functions with respect to the uniform probability measure on $[0,1]^p$.
Consequently, the ambient space of $s_j(\cdot,\cdot, h)$ can be regarded as $L_2^{L_2({\rm unif})}(\tf)$, the space of Hilbertian maps $g:[0,1]^p \to L_2({\rm unif})$ with
\[
\int_{[0,1]^p}\|g(\bx)\|_{L_2({\rm unif})}^2 \tf(\bx)\dx < \infty.	
\]
Let $\cH_+^{L_2({\rm unif})}$ be as defined at \eqref{H-add-sp},  replacing $\R$ with $L_2({\rm unif})$.
Then we consider the following system of equations for weight functions $w_h^{L_2({\rm unif})}\in \cH_+^{L_2({\rm unif})}$ with
$w_h^{L_2({\rm unif})}(\bx)=:w(\bx,\cdot,h)=w_1(x_1,\cdot,h)+\cdots+w_p(x_p,\cdot,h)$:
\begin{equation}\label{wgt-im-sbfeqn}
w_j(x_j,\cdot,h) = s_j(x_j,\cdot,h) -\sum_{k=1, \neq j}^p \int_0^1 w_k(x_k, \cdot,h) \frac{\tf_{j,k}(x_j,x_k)}{\tf_j(x_j)}\, \dxk, \quad 1 \le j \le p.
\end{equation}

According to Theorem 3.4 in \cite{jeon2020additive}, the system of equations (\ref{wgt-im-sbfeqn}) has a unique solution $w_h^{L_2({\rm unif})}$
under the following condition,

\begin{align*}
{\rm(M)} \hspace{0.5cm}\tf_j(x_j)>0, \;\int_0^1\frac{{\tf_{j,k}}^{\,2}(x_j,x_k)}{\tf_k(x_k)}\dxk<\infty \quad &\mbox{for all} \quad x_j \in [0,1] \quad \mbox{and}\quad 1 \le j\neq k \le p,\hspace{1.1cm}\\
\int_{[0,1]^2}\frac{{\tf_{j,k}}^{\,2}(x_j,x_k)}{\tf_j(x_j)\tf_k(x_k)}\dxj \dxk<\infty\quad &\mbox{for all} \quad 1 \le j \neq k \le p.
\end{align*}

This condition holds for sufficiently small $h$ if all $f_j$ are bounded away from zero and the $f_{j,k}$ are continuous
on $[0,1]^2$.
The following proposition provides  a bridge from $m_{+,\R}$ to a new definition for  general metric spaces $\Omega$.

\begin{prop}\label{weight-sum-target}
Assume that condition (M) holds.
Then for real-valued $Y$, the solution $\tm_{+,\R}\in\cH_+^\R$ of \eqref{tilde} and the solution $w(\cdot,\cdot,h)$ of \eqref{wgt-im-sbfeqn}, it holds
that $\tm_{+,\R} = \E\{w(\cdot,\bX,h)Y\}$ and  $\E\{w(\cdot,\bX,h)\} = 1$.
\end{prop}

\begin{proof}
From \eqref{wgt-im-sbfeqn}, we get the following system of equations:
\begin{align}\label{pop-sbfeqn}\begin{split}
&\E\{w_j(x_j,\bX,h)Y\} +\sum_{k=1, \neq j}^p \int_0^1 \E\{w_k(x_k,\bX,h) Y\} \frac{\tf_{j,k}(x_j,x_k)}
{\tf_j(x_j)} \dxk\\
&\qquad = \E\{s_j(x_j,\bX,h)Y\}, \quad 1 \le j \le p.
\end{split}
\end{align}
Comparing the two systems of equations for $\tm_{+,\R}$ and using uniqueness, we obtain
$\tm_{+,\R}=\sum_{j=1}^p \E\{w_j(\cdot,\bX,h)Y\}=\E\{w(\cdot,\bX,h)Y\}$.
To prove the second part, we note that $\E\{s_j(x_j,\bX,h)\}\equiv 1$ from the definition of $s_j$.
Plugging $Y\equiv 1$ into \eqref{pop-sbfeqn}, we see that the set $(\E\{w_1(\cdot,\bX,h)\}, \ldots, \E\{w_p(\cdot,\bX,h)\})
= (1,0,\ldots,0)$, for example, solves \eqref{pop-sbfeqn} with $Y\equiv 1$
using the normalization property $\int \tf_{j,k}(x_j,x_k)\dxk= \tf_j(x_j)$ for all $j \neq k$.
Since the solution of \eqref{wgt-im-sbfeqn} is unique, $\E\{w(\cdot,\bX,h)\}=\sum_{j=1}^p \E\{w_j(\cdot,\bX,h)\}=1$.
\end{proof}

We note that for a general metric space $(\Omega,d)$ which lacks a vector space structure, there is no analogue to \eqref{tilde} since a metric space does not feature algebraic operations.
Nevertheless, Proposition~\ref{weight-sum-target} provides the basis for the proposed additive modeling approach.
Denote by $\Pi(\cdot|\cH_+^\R)$ the projection operator onto $\cH_+^\R$ in $L_2^\R(f)$.
When $Y$ is real-valued, its additive projection is defined as
\begin{equation*}
	m_{+,\R}(\bx) := \Pi\big(\E(Y|\bX=\cdot)\, \big|\,\cH_+^\R\big)(\bx) = \argmin_{a\in\R} \Pi\big(\E\{(Y-a)^2 \mid \bX=\cdot\} \, \big| \, \cH_+^\R \big)(\bx),
\end{equation*}
where the equality follows from the linearity of the projection operator $\Pi(\cdot|\cH_+^\R)$.
Thus $m_{+,\R}(\bx)$ is also the minimizer over $a\in \R$ of the value at $\bx$ for the best additive approximation in $\cH_+^\R$ to $\E((Y-a)^2 | \bX=\cdot)$.

This provides a motivation to define the additive Fr\'echet mean function in a metric space $(\Omega,d)$.
By replacing $(Y-a)^2$ with the squared distance $d^2(Y,a)$, the conditional Fr\'echet variance $\E\{ d^2(Y,a) \mid \bX=\bx\}$ can be approximated by
\begin{equation*}
	M_+(\bx,a):=\Pi\Big(\E(d^2(Y,a)|\bX=\cdot)\big|\cH_+^\R\Big)(\bx).
\end{equation*}
This further suggests approximating the conditional Fr\'echet mean $m_{\Omega}(\bx)$ at \eqref{Frechet-mean} below as follows:
\begin{align}\label{def-m+Omega}
	m_{+,\Omega}(\bx):=\argmin_{a \in \Omega}M_+(\bx,a).
\end{align}
 
 Here the term \emph{additive} refers to the real-valued criterion $M_+(\bx,a)$ for each $a \in \Omega,$ rather than to an algebraic additive decomposition of the $m_{+, \Omega},$ since an additive operation generally is not defined on $(\Omega, d).$ This construction is intrinsic in that it uses only the metric  $d$ and does not require an embedding. 
Note that the smoothed additive projection $\tm_{+,\R}$ provides a close approximation of $m_{+,\R}$ as $h\to0$; under mild conditions, one may show that $\tm_{+,\R}(\bx) - m_{+,\R}(\bx) = O(h^2)$ for each $\bx \in (0,1)^p$.
Applying Proposition~\ref{weight-sum-target} to the real-valued response $d^2(Y,a)$, we further approximate $M_+(\bx,a)$ by
\begin{equation*}
	\tM_+(\bx,a):=\E\{w(\bx,\bX,h) d^2(Y, a)\}.
\end{equation*}
The corresponding smoothed additive regression mean in $(\Omega,d)$ is then defined as
\begin{equation}\label{pop-metric-est}
	\tm_{+,\Omega}(\bx) := \argmin_{a \in \Omega} \tM_+(\bx,a).
\end{equation}
For the approximation of $M_+(\cdot,a)$ by $\tM_+(\cdot,a)$, one may show that
$\tM_+(\bx,a) - M_+(\bx,a) = o(1)$ for each $\bx\in(0,1)^p$ and $a \in \Omega$, provided that $M_+(\cdot,a)$ is continuous at $\bx$ and the standard conditions for kernel regression, such as (B1)--(B3) in \cref{sec:pointwise_asymptotics}, are fulfilled.
This result requires a moment condition for $d^2(Y,a)$ which is automatically satisfied if $\Omega$ is totally bounded.
No other condition on the geometry of $(\Omega,d)$ is needed to obtain the $o(1)$ approximation error.
The approximation can be improved to $O(h^2)$ under further assumptions on the geometry of $(\Omega,d)$, such as (A4) in \cref{sec:pointwise_asymptotics}.
Moreover, it can be strengthened to hold uniformly over $\bx$ and $a\in \Omega$ under a uniform version of condition (A4), such as (U4) in \cref{sec:uniform_asymptotics}.
These extended results follow along the lines of the proofs of Theorems~\ref{bias property} and \ref{uniform bias property} in \cref{sec:pointwise_asymptotics,sec:uniform_asymptotics}, respectively.
Therefore, one may expect that $\tm_{+,\Omega}(\bx)$ in \eqref{pop-metric-est} approximates $m_{+,\Omega}(\bx)$ in \eqref{def-m+Omega}, which is verified using empirical process theory in Section~\ref{sec:asymptotics}.

Notably, Proposition~\ref{weight-sum-target} remains valid for Hilbert-space-valued $Y$, which follows from the  theory developed in \cite{jeon2020additive}. This  shows  that the solution of \eqref{pop-sbfeqn} is also unique for Hilbertian $Y$.
We also note that analogously to other Fr\'echet regression models, additive Fr\'echet regression reduces to $m_{+,\H}=\Pi(\E(Y|\bX=\cdot)|\cH_+^\H)$ when $\Omega=\H$,
where $\cH_+^\H\subset L_2^\H(f)$ denotes a version of $\cH_+^\R$ at \eqref{H-add-sp} constructed by replacing $\R$ with $\H$.
This follows from the observation that
\begin{align}\label{M+real-Y}
M_+(\bx,a)=\Big\|a-\Pi\left(\E(Y|\bX=\cdot)|\cH_+^\H\right)(\bx)\Big\|_\H^2 + c(\bx),
\end{align}
where $\|\cdot\|_\H$ is the norm of $\H$.
The identity at \eqref{M+real-Y} follows from the fact that
\[
\Pi(\mathcal{A}_a(g)|\cH_+^\R)=\mathcal{A}_a(\Pi(g|\cH_+^\H))
\]
for all $g:[0,1]^p \to \H \in L_2^\H(f)$, where $\mathcal{A}_a:\, L_2^\H(f) \to L_2^\R(f)$ is a bounded linear map defined by $\mathcal{A}_a(g)(\cdot)=\langle a,g(\cdot)\rangle_\H$ and $\langle \cdot,\cdot\rangle_\H$ is the inner product $\H$.

In \cref{sec:Add-F-reg-mean-est} below we propose an estimator for $m_{+,\Omega}$ defined at \eqref{def-m+Omega}.
We note that the additive Fr\'echet regression mean $m_{+,\Omega}$ may not coincide with the conditional Fr\'echet regression mean
\begin{align}\label{Frechet-mean}
m_\Omega(\bx):=\argmin_{a \in \Omega}\E(d^2(Y,a)|\bX=\bx), 
\end{align}
where a sufficient condition for their equality  is that $\E(d^2(Y,a)|\bX = \cdot)$ belongs to $\cH_+^\R$ at \eqref{H-add-sp} for every $a\in\Omega$.
This distinction also  arises in ordinary regression: when a linear or additive model is misspecified, its population target need not equal the conditional mean.
This is also the case for the more general Fr\'echet conditional mean and regression, where more flexible modeling approaches that are not subject to the curse of dimensionality such as the proposed additive model  are preferable over less flexible approaches such as global Fr\'echet regression.

\bco

{\color{green} In the literature of Fr\'echet regression, the target of estimation is often not the conditional Fr\'echet mean $m_\Omega$.
This is mainly due to the fact that $\Omega$ in general does not have a vector space structure.
For instance, the global Fr\'echet regression estimator \citep{petersen2019frechet}, say $\widehat m_L(x)$, aims at the minimizer of $\Pi\Big(\E(d^2(Y,a)|\bX=\cdot)\big|\cH_L^\R\Big)(\bx)$
where $\cH_L^\R$ denotes the space of real-valued linear functions in $L_2^\R(f)$ and recall that, according to our notation in the paper, $\Pi(\,\cdot\,|\cH_L^\R)$ is the projection operator onto $\cH_L^R$ in $L_2^\R(f)$.
As another example, the `weak Fr\'echet regression' studied in \cite{bhattacharjee2025nonlinear} estimates the minimizer of
$M_{\rm RKHS}(x,a):=\Pi\Big(\E(d^2(Y,a)|X=\cdot)\big|{\rm RKHS}\Big)(x)$
where ${\rm RKHS}$ represents a reproducing kernel Hilbert space.}

\fi

\begin{exam}\label{example1}
{\rm (Space of networks)}.
Let $(V, E)$ be a network with a set of nodes $V = \{ v_{1}, \ldots, v_{m} \}$ and a set of edge weights $E = \{ w_{ij}: w_{ij} \geq 0,~ i,j = 1,\ldots,m \}$.
Consider the space of networks that are: (i) simple, meaning they contain no self-loops or multi-edges, (ii)  undirected, and
(iii) weighted networks with a fixed number of nodes $m$ and bounded edge weights, i.e., $0\le w_{ij}\le W$ for some $W\ge0$.
Then any network $(V, E)$ in this space can be uniquely determined by its graph Laplacian $L = (l_{ij}),$ where
\begin{equation*}
    l_{ij} =
    \begin{cases}
      -w_{ij}, & i \neq j\\
      \sum_{k \neq i} w_{ik}, & i = j.
    \end{cases}
\end{equation*}
The space of graph Laplacians is given by
\begin{equation*}
    \OL  := \{L \mid L = L^{T},~ L1_{m} = 0_{m},~-W \leq l_{ij} \leq 0 ~~\text{if}~~ i \neq j\},
\end{equation*} where the Frobenius metric is defined as $d_{F}(L_{1}, L_{2}) = (\text{trace}[(L_{1} - L_{2})^{T} (L_{1} - L_{2})])^{1/2}=(\sum_{i=1}^m \sum_{j=1}^m (l_{1,ij} - l_{2,ij})^2)^{1/2}$.
The space    $(\OL, d_{F})$ forms a bounded and convex subset of $\R^{m^{2}}$.
We specify $m_{+,\OL}(\bx)$ and $\tm_{+,\OL}(\bx)$ for the case where $Y$ takes values in $\OL$.
Applying \eqref{M+real-Y} to each entry of $Y$, replacing $\H$ with $\R$, we get
$M_+(\bx,L) = d_F^2\big(\Pi\big(\E(Y|\bX=\cdot)|\cH_+^\R\big)(\bx),L\big) + c(\bx)$ for all $L\in\OL$,
where the projection operator $\Pi(\cdot|\cH_+^\R)$ acts on the matrix function $\E(Y|\bX=\cdot)$ entry-wise
and $c(\cdot)$ does not depend on $L$. Thus in this case $m_{+,\OL}(\bx)$ is the `best' approximation of $\Pi\big(\E(Y|\bX=\cdot)|\cH_+^\R\big)(\bx)$ within $\OL$, i.e.,
\[
m_{+,\OL}(\bx) = \argmin_{L \in \OL} d_F^2\Big(\Pi\big(\E(Y|\bX=\cdot)|\cH_+^\R\big)(\bx), \, L\Big).
\]
Similarly, we have $\tm_{+,\OL}(\bx) = \argmin\limits_{L \in \OL} d_F^2\big(\E(w(\bx,\bX,h)Y), \, L\big)$.
These minimization problems can be solved by  quadratic programming  \citep{zhou2022network}.
\end{exam}

\begin{exam}\label{example2}
{(\rm $2$-Wasserstein space of univariate distributions)}.
Let $\mathcal{P}(\mathcal{D})$ denote the space of probability measures on a compact domain $\mathcal{D \subset \R}$.
For the space of probability measures with finite second moments
$ \Omega_{W} :=  \left\{ \nu \in \mathcal{P}(\mathcal{D}): \int_{\mathcal{D}} \lVert s \rVert^{2} d\nu(s) < \infty \right\}$,
the $2$-Wasserstein metric $d_{W}$ between any two probability measures $\nu_{1}, \nu_{2} \in \Omega_{W}$ is
\begin{equation*}
	d_{W}(\nu_1,\nu_2) := \left( \int_{0}^{1} \left\{F_{\nu_{1}}^{-1}(s) -  F_{\nu_{2}}^{-1}(s)\right\}^2 ds \right)^{1/2},
\end{equation*}
where $F_{\nu_{1}}^{-1}(\cdot),$ $F_{\nu_{2}}^{-1}(\cdot)$ are the quantile functions of $\nu_{1}$ and $\nu_{2},$ respectively.
The space of univariate probability distributions equipped with the $2$-Wasserstein distance $(\Omega_{W}, d_{W})$ is both complete and separable.
To specify $m_{+,\Omega_W}(\bx)$ and $\tm_{+,\Omega_W}(\bx)$ in this case, we note that the projection operator $\Pi(\cdot|\cH_+^\R)$ is linear so that
\[
M_+(\bx,\nu)=\int_0^1\Big(\Pi\big(\E({F_{Y}^{-1}(s)} | \bX=\cdot) |\cH_+^\R \big)(\bx)-F_{\nu}^{-1}(s)\Big)^2 ds + c(\bx), \quad \nu \in \Omega_W
\]
for some $c(\cdot)$ that is independent of $\nu$. Thus
$m_{+,\Omega_W}(\bx)=\argmin\limits_{\nu \in \Omega_W}\int_0^1 \big[ \Pi\big(\E({F_{Y}^{-1}(s)} | \bX=\cdot) |\cH_+^\R \big)(\bx)-F_{\nu}^{-1}(s) \big]^2 ds$,
and similarly, $\tm_{+,\Omega_W}(\bx) = \argmin\limits_{\nu \in \Omega_W} \int_0^1 \big[ \E\{w(\bx,\bX,h)F_Y^{-1}(s)\} - F_\nu^{-1}(s) \big]^2 ds$.
These minimization problems can again be solved by quadratic programming  \citep{petersen2019frechet}.
\end{exam}

\begin{exam}\label{example3}
{(\rm Space of compositional and spherical data)}.
The $(p-1)$-dimensional simplex where compositional data reside is given by
\begin{equation*}
    \Omega_{C} := \left\{ (a_1, \ldots, a_{p}) \in \R^{p} \mid a_j \geq 0, ~j = 1,\ldots,p,~\sum_{i=1}^{p} a_{i} = 1,  \right\}.
\end{equation*}
Consider the component-wise square root transformation $(a_1, \ldots, a_p) \mapsto (\sqrt{a_1}, \ldots, \sqrt{a_{p}}),$ which maps $\Omega_{C}$ to the unit sphere
\begin{equation*}
    \mathbb{S}^{p-1}_{+} := \left\{\bb = (b_1, \ldots, b_p) \mid b_1^2 +\cdots+b_p^2 = 1,~ b_j \geq 0,~ j=1,\ldots, p \right\},
\end{equation*} equipped with the geodesic Riemannian distance  $d_g(\bb_1 , \bb_2) = \arccos(\bb_1^{T}\bb_2),$ for $\bb_1, \bb_2 \in \mathbb{S}^{p-1}_{+}$ \citep{scea:14}.
For spherical data with the Riemannian metric, empirical minimizers are obtained with  gradient descent \citep{groisser2004newton,pennec2013exponential}.
\end{exam}

\subsection{Additive \F regression estimator}\label{sec:Add-F-reg-mean-est}

Assume we observe $n$ independent and identically distributed (i.i.d.) copies $(\bX_1,Y_1), \ldots, (\bX_n,Y_n)$  of $(\bX,Y)$. Denote
the $j$th entry of $\bX_i$ by  $X_{ij}$,  i.e., $\bX_i\equiv(X_{i1},\ldots,X_{ip})$.
The additive Fr\'echet regression mean $m_{+,\Omega}$ can be estimated by an empirical version of $\tm_{+,\Omega}$ \eqref{pop-metric-est}.
Let $\hw_h^{L_2({\rm unif})}$ be the version of $w_h^{L_2({\rm unif})}$ that is obtained by replacing $s_j, \, \tf_j$ and $\tf_{j,k}$ at \eqref{wgt-im-sbfeqn}
with $\hs_j,\,\hf_j$ and $\hf_{j,k}$, respectively, where
\begin{align}\label{ker-den-est}
\begin{split}
  \hs_j(\bx,\bu,h)&\equiv \hs_j(x_j,\bu,h)\equiv \hs_j(x_j,u_j,h):=\frac{K_h(x_j,u_j)}{\hf_j(x_j)}, \quad 1 \le j \le p,\\
\hf_j(x_j)&:=\frac 1 n \sum_{i=1}^n K_h(x_j,X_{ij}),\quad 1 \le j \le p,\\
\hf_{j,k}(x_j,x_k)&:=\frac 1 n \sum_{i=1}^n K_h(x_j,X_{ij})K_h(x_k,X_{ik}), \quad 1 \le j \neq k \le p.
\end{split}
\end{align}
According to Theorem 3.4 in \cite{jeon2020additive}, $\hw_h^{L_2({\rm unif})}$ is uniquely determined under the following conditions:
\begin{align*}
\hf_j(x_j)>0, \;\int_0^1\frac{{\hf_{j,k}}^{\,2}(x_j,x_k)}{\hf_k(x_k)}\dxk<\infty \quad &\mbox{for all} \quad x_j \in [0,1] \quad \mbox{and}\quad 1 \le j \neq k\le p,\\
\int_{[0,1]^2}\frac{{\hf_{j,k}}^{\,2}(x_j,x_k)}{\hf_j(x_j)\hf_k(x_k)}\dxj \dxk<\infty\quad &\mbox{for all} \quad 1 \le j \neq k \le p.
\end{align*}
The above conditions hold with probability tending to one if all $f_j$ are bounded away from zero and infinity and the $f_{j,k}$ are continuous
on $[0,1]^2$.

Consider weight functions  $\hw: [0,1]^p \times [0,1]^p \rightarrow \R$ given  by
\begin{equation}\label{wgt-estimate}
    \hw(\bx,\cdot,h):=\hw_h^{L_2({\rm unif})}(\bx).
\end{equation}
By similar arguments  as in the proof of Proposition~\ref{weight-sum-target} it holds  that
\begin{align*}
n^{-1}\sum_{i=1}^n \hw(\cdot,\bX_i,h) =1.
\end{align*}
An empirical version of $\tM_+(\cdot,a)$ is $\hM_+(\cdot,a):=n^{-1}\sum_{i=1}^n \hw(\cdot,\bX_i,h)d^2(Y_i,a)$,
leading to
\begin{equation}\label{def-m+Omega-sample}
    \hatm_{+,\Omega}:=\argmin_{a \in \Omega}\hM_+(\cdot,a).
\end{equation}
We observe that $\hatm_{+,\Omega}$ reduces to $n^{-1}\sum_{i=1}^n \hw(\bx,\bX_i,h) Y_i$ when $\Omega=\H$, which equals exactly the Hilbertian additive regression estimator studied in \cite{jeon2020additive}.
Indeed, when $\Omega=\H$,
\begin{align*}
\hM_+(\bx,a)&=n^{-1}\sum_{i=1}^n \hw(\bx,\bX_i,h)\|Y_i-a\|_\H^2\\
&=\Big\|a-n^{-1}\sum_{i=1}^n \hw(\bx,\bX_i,h) Y_i \Big\|_\H^2 + \mbox{(irrelevant term)},
\end{align*}
where we have used $n^{-1}\sum_{i=1}^n \hw(\cdot,\bX_i,h) =1$.
Regarding implementation details for the estimator $\hatm_{+,\Omega}$ at \eqref{def-m+Omega-sample}, we refer to Section~\ref{sec:net_nu_imp}.

\section{Asymptotic analysis}\label{sec:asymptotics}

\subsection{Pointwise convergence rates}\label{sec:pointwise_asymptotics}

For all theoretical results,  we assume throughout that $(\Omega,d)$ is totally bounded, which is a common assumption in the Fr\'echet regression literature \citep{petersen2019frechet,chen2022uniform,ghosal2023frechet,bhattacharjee2023single}.
Throughout this subsection, we fix a point of interest $\bx\equiv (x_1, \ldots, x_p) \in (0,1)^p$ and aim to estimate $m_{+,\Omega}(\bx)$.
Let $\partial_j^k:C^k([0,1]^p)\to C^0([0,1]^p)$ for $1 \le j \le p$ and $k=1,2$ be the differential operators such that $(\partial_j^k g)(\bz):=(\partial^k/\partial z_j^k)g(\bz)$.
Note that $(\partial_j^k g)(\bz)\equiv(\partial_j^k g_j)(z_j)$ for any $g \in \cH_+^\R$ and decompositions $g=g_1+\cdots +g_p$ with $g_j \in \cH_j^\R$.
For a subset $\cS$ of a metric space $(\cM, d_{\cM})$, denote by $N(\epsilon,\cS,d_\cM)$ the $\epsilon$-covering number of $\cS$.
For sequences of positive numbers $a_n$ and $b_n$, we write $a_n \lesssim b_n$ if there exists a constant $0<C<\infty$ such that $a_n \le C\,b_n$ for all $n$.
For $\delta>0$, let $B_\delta(a_0):=\{a \in \Omega: d(a,a_0)< \delta\}$.

In our theoretical development we use various concentration inequalities in empirical process theory.
For this we need some technical assumptions on the curvatures of $M_+$ and $\tM_+$ as well as on the entropy of $\Omega$.
The following assumptions will be used for the asymptotic analysis of the additive Fr\'echet regression estimator $\hatm_{+,\Omega}$ defined at \eqref{def-m+Omega-sample}.
These assumptions are satisfied by commonly encountered non-Euclidean spaces, including the spaces discussed in the examples in \cref{sec:Add-F-reg-mean}.

\begin{enumerate}
\item[(A1)] The function $M_+(\bx,\cdot)$ is well defined for each $a\in\Omega$ and has a unique minimizer $m_{+,\Omega}(\bx)$ and
for any $\epsilon>0$ it holds that \newline 
$\inf_{a\in\Omega:d(a,m_{+,\Omega}(\bx))>\epsilon}\left\{M_+(\bx,a)-M_+(\bx,m_{+,\Omega}(\bx))\right\}>0$.
\item[(A2)] There exist $\eta>0, \, C_1>0$ and $\alpha>1$ such  that for all $a \in \Omega$ with $d(a, m_{+,\Omega}(\bx))<\eta$,
\begin{align*}
M_+(\bx,a)-M_+(\bx,m_{+,\Omega}(\bx)) \ge C_1 d(a, m_{+,\Omega}(\bx))^{\alpha}.
\end{align*}

\item[(A3)] There exist $\gamma>0, \, C_2>0,\,\beta>1$ and $h_0>0$ such that for all $h\le h_0$ and for all $a \in \Omega$ with $d(a, \tm_{+,\Omega}(\bx))<\gamma$,
\begin{align*}
\tM_+(\bx,a)-\tM_+(\bx,\tm_{+,\Omega}(\bx)) &\ge C_2 d(a, \tm_{+,\Omega}(\bx))^{\beta}.
\end{align*}

\item[(A4)] For each $a\in \Omega$, $M_+(\cdot,a)$ is twice partially differentiable and there exist $\kappa>0$ and $C_3>0$ such that for all $a \in \Omega$ with $d(a, m_{+,\Omega}(\bx))<\kappa$,
\[
\max_{1\le j \le p}\sup_{\bz \in [0,1]^p}|(\partial_j^k M_+)(\bz,a)-(\partial_j^k M_+)(\bz,m_{+,\Omega}(\bx))| \le C_3\,d(a, m_{+,\Omega}(\bx)), \quad k=1,2.
\]

\item[(A5)] There exists $\delta_0>0$ such that
\begin{align*}
\sup_{0<\delta <\delta_0}\int_0^1 \sqrt{1+\log N\big(t\delta, B_\delta(\tm_{+,\Omega}(\bx)),d\big)}\,{\rm d}t =O(1).
\end{align*}
\end{enumerate}
Here (A1)  ensures the uniqueness of $m_{+,\Omega}(\bx)$ and also is needed to establish the consistency of $\hatm_{+,\Omega}(\bx)$; it implies that  small $\sup_{a\in\Omega}|\hM_+(\bx,a) - M_+(\bx,a)|$ entails small $d(\hatm_{+,\Omega}(\bx),m_{+,\Omega}(\bx))$.
Assumptions (A2) and (A3) are used  to derive the convergence rates of $d(\tm_{+,\Omega}(\bx),m_{+,\Omega}(\bx))$ and $d(\hatm_{+,\Omega}(\bx),\tm_{+,\Omega}(\bx))$, in Theorems~\ref{bias property} and \ref{var property}.
We also note that the constants in (A1)--(A5) implicitly depend on the number of covariates $p$.

If the response space $(\Omega, d)$ is a complete geodesic space with nonpositive curvature (i.e., a  Hadamard or complete CAT(0) space) and the conditional \F function $\E (d^2(Y, a) \mid \bX = \bs{\cdot})$ has an additive structure in $\mathcal{H}^{\R}_{+}$, $M_{+}(\bx, a)$ has a unique minimizer, ensuring the identifiability in Assumption (A1) \citep{sturm2003probability}, and (A2) also holds with quadratic curvature parameters $\alpha = 2$ \citep{dubey2022modeling}. 
Hadamard spaces form a broad class, including graph Laplacians under the Frobenius metric (Example 1) and univariate distributions under the 2-Wasserstein distance (Example 2); see also  \cite{song2025inference} for further examples.
For positively curved spaces such as the  sphere $\mathbb{S}^{p-1}$ with the geodesic distance, uniqueness of the \F regression in (A1) may fail in general.
However, in restricted settings such as the positive orthant $\mathbb{S}_{+}^{p-1}$ (Example 3), the required conditions can still be established \citep{afsari2011riemannian, lin2021total}.
More generally, in positively curved spaces, whether these conditions hold depends not only on the geometry of the space but also on the underlying distribution of $(\bX, Y)\sim F$.

Similarly, (A3) is satisfied by graph Laplacians and univariate distributions in Examples 1 and 2.
Note that (A3) involves possibly negative smooth backfitting weights $w$ solving \eqref{wgt-im-sbfeqn}, and hence its validity requires additional conditions on both the geometry of $\Omega$ and the distribution of $(\bX,Y)$; see Section S.1.1 for further discussion.
(A5) is needed to analyze $\tM_+(\bx,a) - M_+(\bx,a)$ and $\hM_+(\bx,a) - \tM_+(\bx,a)$ for $a\in\Omega$, while (A4) is a mild distributional assumption, weaker than (L1) in \cite{petersen2019frechet}, and is required to establish the bias expansion in Theorem~\ref{bias property}.
Note that our theoretical results do not require the true additive structure $M_{+}(\bs{\cdot}, a) = \E (d^2(Y, a) \mid \bX = \bs{\cdot}) \in \mathcal{H}^{\R}_{+}$; the additive structure is used only to provide geometric intuition and concrete sufficient conditions under which the assumptions can be verified.

The following additional  assumptions are required to investigate the stochastic part of the estimator $\hatm_{+,\Omega}(\bx)$; these are
 a subset of conditions  typically required in the standard theory of smooth backfitting for Hilbert space-valued responses.

\begin{enumerate}
\item[(B1)] The baseline kernel function $K$ is nonnegative, symmetric, bounded, Lipschitz continuous,
supported on $[-1,1]$ and satisfies $\int K=1$.
\item[(B2)] All marginal densities $f_j$ are bounded away from zero on $[0,1]$,
and all joint densities $f_{j,k}$ are continuous on $[0,1]^2$.
\item[(B3)] The bandwidth satisfies  $h \to 0$ and $(\log n)^2/(nh^2) \lesssim 1$ as $n \to \infty$.
\end{enumerate}

\begin{theorem}\label{consistency}
Assume (A1) and (B1)--(B3).
Also, assume that, for each $a\in\Omega$, $M_+(\cdot,a)$ is continuous at $\bx$.
Then $d(\hatm_{+,\Omega}(\bx), m_{+,\Omega}(\bx))=o_p(1)$.
\end{theorem}

The estimation error of $\hatm_{+,\Omega}(\bx)$ in terms of $d(\hatm_{+,\Omega}(\bx), m_{+,\Omega}(\bx))$ is bounded by
the sum of the deterministic part $d(\tm_{+,\Omega}(\bx),m_{+,\Omega}(\bx))$ and the stochastic part $d(\hatm_{+,\Omega}(\bx), \tm_{+,\Omega}(\bx))$.
The next two theorems give the magnitudes of these two errors.

\begin{theorem}\label{bias property}
Assume (A1), (A2), (A4), (B1) and (B2). Then $d(\tm_{+,\Omega}(\bx), m_{+,\Omega}(\bx))=O(h^{\frac{2}{\alpha-1}})$
for sufficiently small $h>0$.
\end{theorem}

\begin{theorem}\label{var property}
Assume (A1), (A3), (A5) and (B1)-(B3). Then
\[
d(\hatm_{+,\Omega}(\bx), \tm_{+,\Omega}(\bx))=O_p\Big((nh)^{-\frac{1}{2(\beta-1)}}\Big).
\]
\end{theorem}

We give a heuristic account of the theoretical challenge involved in the proof of Theorem~\ref{var property}.
A central challenge  is to establish
\begin{align}\label{key-innovation}\begin{split}
&\sup_{a\in B_\delta(\tm_{+,\Omega}(\bx))}\Big|n^{-1}\sum_{i=1}^n \big\{\hw(\bx,\bX_i,h)-w(\bx,\bX_i,h)\big\}\big\{d^2(Y_i,a)-d^2(Y_i,\tm_{+,\Omega}(\bx))\big\}\Big|\\
&\qquad = \delta \cdot O_p(n^{-1/2}h^{-1/2})
\end{split}
\end{align}
uniformly for all $0<\delta<\delta_0$, where $\delta_0$ is the constant in the assumption (A5) and $O_p(\cdot)$ does not depend on $\delta.$
This requires using various projection operators and associated linear operators for complex stochastic functions taking values in $\cH_+^\R$.
Let $\hpi_j: L_2^\R(\hf)\to \cH_j^\R$ be the projection operators
\[
\hpi_j(g)(z_j):=\int g(\bz)\frac{\hf(\bz)}{\hf_j(z_j)}\,{\rm d}\bz_{-j}, \quad g \in L_2^\R(\hf), \quad 1 \le j \le p,
\]
where $\hf(\bz) := n^{-1}\sum_{i=1}^n \prod_{j=1}^p K_h(z_j,X_{ij})$.
Define $\tpi_j:L_2^\R(\tf)\to \cH_j^\R$ by replacing $\hf$ and $\hf_j$ with $\tf$ and $\tf_j$, respectively, in the above definition.
For $\bz \in [0,1]^p$ and $1 \le j \le p$, let
\[
\nu_j(\bz,a)\equiv\nu_j(z_j,a):=n^{-1}\sum_{i=1}^n s_j(z_j,X_{ij},h) \big\{d^2(Y_i,a)-d^2(Y_i,\tm_{+,\Omega}(\bx))\big\}.
\]
Here and in expressions derived from $\nu_j$, we suppress the dependence on $\bx$ for simplicity.

Define a linear operator $\hcT: \cH_+^\R \to \cH_+^\R$ by $\hcT:=(\rI-\hpi_p)\circ\cdots \circ(\rI-\hpi_1)$, where $\rI: \cH_+^\R \to \cH_+^\R$ denotes the identity map, and let 
\begin{align*}
\htau(\bz,a):=\nu_p(\bz,a) + (\rI-\hpi_p)(\nu_{p-1}(\cdot,a))(\bz)+ \cdots + (\rI-\hpi_p)\circ\cdots \circ(\rI-\hpi_2)(\nu_1(\cdot,a))(\bz).
\end{align*}
Also, define  $\tcT: \cH_+^\R \to \cH_+^\R$ and $\ttau(\cdot,a)$ by replacing $\hpi_k$ with $\tpi_k$ for $1 \le k \le p$ in the above definitions of $\hcT$ and $\htau(\cdot,a)$. A 
key component in the proof of \eqref{key-innovation} is to verify that the $L_2$ and $L_\infty$ norms of various terms in the decompositions of
\[
\htau(\cdot,a)-\ttau(\cdot,a), \quad \tcT\Big(\sum_{r=0}^\infty \tcT^r\left(\htau(\cdot,a)-\ttau(\cdot,a)\right)\Big), \quad (\hcT-\tcT)\Big(\sum_{r=0}^\infty \tcT^r\left(\htau(\cdot,a)\right)\Big)
\]
are of magnitude $\delta \cdot O_p(n^{-1/2}h^{-1/2})$ uniformly over $a \in B_\delta(\tm_{+,\Omega}(\bx))$ and over $0<\delta<\delta_0$.
We refer to the proofs of Lemmas~S.4 and S.5 for details.

The uniformity over $\delta$ at \eqref{key-innovation} is necessary,  as for  the proof of Theorem~\ref{var property} we require
an upper bound of the left-hand side of \eqref{key-innovation} that is uniform over a discrete set of $\delta$
whose elements are in the range $0<\delta<\delta_0$ and whose cardinality diverges as the sample size increases.
The verification of these claims is quite challenging and  requires invoking  various empirical processes indexed by function classes
that arise  when applying the projections $\hpi_j$ or $\tpi_j$ for the  $\nu_k$ for $j \neq k$,
which is absent in the existing theory of smooth backfitting.
The technical arguments that are required to ensure the uniformity over $\delta$ as well as over the ball $B_\delta(\tm_{+,\Omega}(\bx))$
add additional complexity.
We note that the standard theory of additive regression for Hilbert-space-valued responses requires only that $\|\hpi_j-\tpi_j\|_{L_2(f)}=o_p(1)$
and $\|\hcT - \tcT\|_{\rm op}=o_p(1)$ \citep{jeon2020additive}. These bounds do not suffice for the present analysis, as  the contributions of $(\hpi_j -\tpi_j)$ and
$(\hat\cT - \tilde\cT)$ are no longer negligible, but constitute the leading terms for  $d(\hatm_{+,\Omega}(\bx), \tm_{+,\Omega}(\bx))$.

Combining Theorems~\ref{bias property} and \ref{var property}, we obtain
\begin{align}\label{MSE property}
d(\hatm_{+,\Omega}(\bx), m_{+,\Omega}(\bx))=O_p\Big(h^{\frac{2}{\alpha-1}}+(nh)^{-\frac{1}{2(\beta-1)}}\Big).
\end{align}
The rate is the same as that of the local Fr\'echet regression estimator with a covariate $X$ taking values in $\R$
\citep{petersen2019frechet}, which means that our approach is free from the curse of dimensionality.
When $\alpha=\beta=2$ in the assumptions (A2) and (A3), the rate in \eqref{MSE property} reduces to $h^2+n^{-1/2}h^{-1/2}$, which is the typical univariate rate for real-valued or Hilbert-space-valued
responses \citep{mammen1999existence,jeon2020additive}.
We note that the assumptions with $\alpha=\beta=2$ hold for the spaces discussed in Examples~\ref{example1} and \ref{example3} \citep{petersen2019frechet,kurisu2024geodesic} and
they are also satisfied for finite-dimensional Riemannian manifolds.
For Example~\ref{example2}, we note that the entropy calculation in  \cite{petersen2019frechet} for this specific case contains an error
and additional conditions are needed to satisfy assumption (A5)  for $\alpha=\beta=2$  for the 2-Wasserstein space.
A sufficient condition is that $\Omega_W$ is a  closed, convex, finite-dimensional subset of the Wasserstein space.
This is satisfied for example for a set that consists of convex combinations of a finite number of quantile functions.
This also applies to assumption (U5) in the following section.
Substituting optimal bandwidths  $h_{\rm opt} \asymp n^{-1/5}$ then leads to  $d(\hatm_{+,\Omega}(\bx), m_{+,\Omega}(\bx))=O_p(n^{-2/5})$, which is known to be minimax optimal for the class of regression functions
with bounded second derivatives when $Y$ is real-valued.

Considering multivariate versions of Theorems 3 and 4 in \cite{petersen2019frechet} for the `full-dimensional' local Fr\'echet regression estimator $\hatm_\Omega$
defined by
\begin{equation}\label{local_frechet_estimator}
\hatm_\Omega(\bx):=\argmin_{a \in \Omega} \left(n^{-1}\sum_{i=1}^n \bigg\{\prod_{j=1}^p \frac{1}{h}K\bigg(\frac{x_j-X_{ij}}{h}\bigg)\bigg\}d^2(Y_i,a)\right)
\end{equation}
the target of this full-dimensional unstructured estimator is not the structured Fr\'echet regression mean $m_{+,\Omega}(\bx)$ studied in this paper, but is $m_\Omega(\bx)$ defined at \eqref{Frechet-mean}.
Nevertheless, it is of interest to compare the error rate of our additive estimator $\hatm_{+,\Omega}(\bx)$ with that of the unstructured $\hatm_\Omega(\bx)$.
Under suitable assumptions, following the technical developments in \cite{petersen2019frechet}, it is tedious but straightforward to show that
\begin{align}\label{pointwise MSE local_frechet}
d(\hatm_\Omega(\bx), m_\Omega(\bx))=O_p\Big(h^{\frac{2}{\alpha-1}}+(nh^p)^{-\frac{1}{2(\beta-1)}}\Big).
\end{align}
Comparing this with \eqref{MSE property}, we find that the stochastic error of size $(nh^p)^{-\frac{1}{2(\beta-1)}}$ deteriorates for large $p$, while 
for fixed $p$ the proposed estimator $\hatm_{+,\Omega}(\bx)$ attains the univariate rate.

\subsection{Uniform convergence rates}\label{sec:uniform_asymptotics}

To obtain  uniform rather than pointwise rates of convergence of $\hatm_{+,\Omega}$, we require a uniform version of the assumption set (A) in \cref{sec:pointwise_asymptotics}.

\begin{enumerate}
\item[(U1)] $M_+(\bx,a)$ is well defined for all $\bx\in[0,1]^p$ and $a\in\Omega$.
Also, $\hatm_{+,\Omega}(\bx)$ and $m_{+,\Omega}(\bx)$ exist uniquely for all $\bx\in[0,1]^p$.
Furthermore, for any $\epsilon>0$, it holds that
\begin{align*}
\inf_{\bx\in[0,1]^p}\inf_{a\in\Omega:d(a,m_{+,\Omega}(\bx))>\epsilon}\{M_+(\bx,a)-M_+(\bx,m_{+,\Omega}(\bx))\} & > 0, \\
\liminf_{h\to0}\inf_{\bx\in[0,1]^p}\inf_{a\in\Omega:d(a,\tm_{+,\Omega}(\bx))>\epsilon}\{\tM_+(\bx,a)-\tM_+(\bx,\tm_{+,\Omega}(\bx))\} & > 0,	
\end{align*}
and there exists $c(\epsilon)>0$ such that
\begin{equation*}
\lim_{n\to\infty}\P\Big( \inf_{\bx\in[0,1]^p}\inf_{a\in\Omega:d(a,\hatm_{+,\Omega}(\bx))>\epsilon}\{\hM_+(\bx,a)-\hM_+(\bx,\hatm_{+,\Omega}(\bx))\}>c(\epsilon) \Big) = 1.
\end{equation*}
\item[(U2)] There exist constants $\eta_U>0, \, C_{1,U}>0$ and $\alpha_U>1$ such that, for each $\bx\in[0,1]^p$,
\begin{align*}
M_+(\bx,a)-M_+(\bx,m_{+,\Omega}(\bx)) \ge C_{1,U} d(a, m_{+,\Omega}(\bx))^{\alpha_U}
\end{align*}
for all $a\in\Omega$ with $d(a,m_{+,\Omega}(\bx))<\eta_U$.
\item[(U3)] There exist constants $\gamma_U>0, \, C_{2,U}>0,\,\beta_U>1$ and $h_{0,U}>0$ such that, for each $\bx\in[0,1]^p$,
\begin{align*}
\tM_+(\bx,a)-\tM_+(\bx,\tm_{+,\Omega}(\bx)) \ge C_{2,U} d(a, \tm_{+,\Omega}(\bx))^{\beta_U}
\end{align*}
for all $a\in\Omega$ with $d(a,\tm_{+,\Omega}(\bx))<\gamma_U$ and for all $h\le h_{0,U}$.
\item[(U4)] $M_+(\cdot,a)$ has bounded second-order partial derivatives uniformly over $\bx\in[0,1]^p$ and $a\in\Omega$.
In addition, there exist $\kappa_U>0$ and $C_{3,U}>0$ such that
\[
\max_{1\le j \le p}\sup_{\bz \in [0,1]^p}|(\partial_j^k M_+)(\bz,a)-(\partial_j^k M_+)(\bz,m_{+,\Omega}(\bx))|\le C_{3,U}\,d(a, m_{+,\Omega}(\bx)), \quad k=1,2,
\]
for all $a \in \Omega$ with $d(a, m_{+,\Omega}(\bx))<\kappa_U$ and for all $\bx\in[0,1]^p$.
\item[(U5)] There exists a constant $\delta_{0,U}>0$ such that
\begin{align*}
\sup_{0<\delta <\delta_{0,U}}\int_0^1 \sup_{\bx\in[0,1]^p}\sqrt{1+\log N\big(t\delta, B_\delta(\tm_{+,\Omega}(\bx)),d\big)}\,{\rm d}t =O(1).
\end{align*}
\end{enumerate}

Assumptions (U1)--(U5) are the uniform versions of (A1)--(A5), and are required to show that Theorems~\ref{consistency}--\ref{var property} hold uniformly for
$\bx \in [0,1]^p$.
Comparing (U1) and (A1), in particular, the additional assumptions on $\tM_+$ and $\hM_+$ in (U1) are to ensure the asymptotic equicontinuity of $\tm_{+,\Omega}(\cdot)$ and $\hatm_{+,\Omega}(\cdot)$, respectively.
As for the assumption set (B), we  replace (B1) by (B1$'$) given below.
\begin{enumerate}
\item[(B1$'$)] The baseline kernel function $K$ satisfies (B1) and its first derivative is Lipschitz continuous.
\end{enumerate}
The Lipschitz continuity of the first derivative of $K$ in (B1$'$) is required to derive some asymptotic bounds for the partial derivatives of $w_j(x_j,\cdot,h)$ with respect to $x_j$ in the system of equations at \eqref{wgt-im-sbfeqn}.
The bounds in turn imply the Lipschitz continuity of $\tm_{+,\Omega}$, which is instrumental in establishing the uniform convergence rate over $\bx\in[0,1]^p$ via empirical process theory.

The following results are uniform versions of Theorems~\ref{consistency}--\ref{var property}.

\begin{theorem}\label{uniform consistency}
Assume (U1), (U4), (B1$'$), (B2) and (B3). Then
\begin{equation*}
\sup_{\bx \in [0,1]^p}d(\hatm_{+,\Omega}(\bx), m_{+,\Omega}(\bx))=o_p(1).
\end{equation*}
\end{theorem}

\begin{theorem}\label{uniform bias property}
Assume (U1), (U2), (U4), (B1$'$) and (B2). For sufficiently small $h>0$,
\begin{align*}
\sup_{\bx \in [2h,1-2h]^p}d(\tm_{+,\Omega}(\bx),m_{+,\Omega}(\bx)) &= O(h^{\frac{2}{\alpha_U-1}}), \text{ and}\\
\sup_{\bx \in [0,1]^p}d(\tm_{+,\Omega}(\bx),m_{+,\Omega}(\bx)) &= O(h^{\frac{1}{\alpha_U-1}}),
\end{align*}
where $\alpha_U$ is the constant in Assumption (U2).
\end{theorem}

We note that the difference between the rates for the interior $[2h,1-2h]^p$ and for the boundary $[0,1]^p\setminus [2h,1-2h]^p$
in Theorem~\ref{uniform bias property} is typical when using  Nadaraya-Watson type kernel estimators  that we adopt for the smoothing steps  in this paper.

\begin{theorem}\label{uniform var property}
Assume (U1), (U3), (U5), (B1$'$), (B2) and (B3). Then
\[
\sup_{\bx \in [0,1]^p}d(\hatm_{+,\Omega}(\bx), \tm_{+,\Omega}(\bx))=O_p\Big(\Big(\frac{\log n}{nh}\Big)^{\frac{1}{2(\beta_U-1)}}\Big),
\]
where $\beta_U$ is the constant that appears in (U3).
\end{theorem}
A consequence of  Theorems~\ref{uniform bias property} and \ref{uniform var property} is that 
\begin{align}\label{unif-MSE property}\begin{split}
\sup_{\bx \in [2h,1-2h]^p}d(\hatm_{+,\Omega}(\bx), m_{+,\Omega}(\bx)) &= O_p\bigg(h^{\frac{2}{\alpha_U-1}} +  \bigg(\frac{\log n}{nh}\bigg)^{\frac{1}{2(\beta_U-1)}}\bigg), \text{ and}\\
\sup_{\bx \in [0,1]^p}d(\hatm_{+,\Omega}(\bx), m_{+,\Omega}(\bx)) &= O_p\bigg(h^{\frac{1}{\alpha_U-1}} +  \bigg(\frac{\log n}{nh}\bigg)^{\frac{1}{2(\beta_U-1)}}\bigg).
\end{split}
\end{align}
For fixed $p$, the exponents of the convergence rates at \eqref{unif-MSE property} do not involve $p$, although the constants appearing in (U1)--(U5) may depend on $p$.
Thus the proposed additive model avoids the curse of dimensionality with respect to both pointwise and uniform convergence.

\section{Simulations}\label{sec:simul}
\subsection{Implementation details}\label{sec:net_nu_imp}

We assess the finite sample properties of the proposed estimator $\hatm_{+.\Omega}$ in terms of estimating the underlying Fr\'echet regression function $m_\Omega$ defined
at \eqref{Frechet-mean}. In scenarios where $m_\Omega$ admits the {\it intrinsic} additive structure of  \cref{sec:Add-F-reg-mean}, i.e., $m_\Omega=m_{+,\Omega}$, it is expected
that $\hatm_{+,\Omega}$ performs well in view of  the theoretical guarantees demonstrated in \cref{sec:asymptotics}, as  $m_{+,\Omega}$ is the target of $\hatm_{+,\Omega}$.
We investigate here  how is reflected in the finite sample case  and also study the finite sample  performance when $m_\Omega$ deviates from an additive structure. 


To compute the estimator $\hatm_{+,\Omega},$ we solve the minimization problem in \eqref{def-m+Omega-sample}, where the optimization algorithm depends on the geometry of the response space $(\Omega, d)$.
For network-valued responses $(\OL, d_F)$ represented via graph Laplacians, the minimization reduces to a quadratic program over a closed and convex space (Section 5.1 of \ci{zhou2022network}), while
for distributional responses $(\Omega_{W}, d_W)$ equipped with the 2-Wasserstein distance, the objective similarly takes the form of a quadratic program (Section 6.1 of \ci{petersen2019frechet}).
These optimization problems can be efficiently solved using the OSQP solver \citep{stellato2020osqp}, which we implement via the osqp R package.

For a general nonlinear metric space, however, the presence of negative weights 
may yield nonconvex objective functions and lead to non-unique minimizers, in which case gradient-based algorithms ensure only local algorithmic convergence.
The uniqueness of the \F mean depends not only on the geometry of the metric space but also on the distribution of $(\bX,Y)$.
Establishing algorithmic convergence may also require additional assumptions on the initialization and the specific algorithm employed, and optimization in general metric spaces requires a case-by-case analysis.
Further discussion of  the effect of negative weights and additional implementation details, including initialization, hyperparameter selection, runtime and bandwidth choice, are provided in the Supplementary Material Section S.6.
Graphical illustrations of the weights $\hat{w}$ can be found in Supplementary Material Section S.7.
Our  simulations encompass network-valued and distribution-valued responses, with details reported in Sections \ref{sec:net_simul} and \ref{sec:den_simul}, respectively.

\subsection{Networks as responses via graph Laplacians}\label{sec:net_simul}

Consider an $m \times m$ graph Laplacian-valued response $L$ in the space of networks $\OL$,  equipped with the Frobenius metric $d_F$, described in Example 1 of Section \ref{sec:Add-F-reg-mean}
with predictors $\bX$ residing in $\R^{3}$.  For any symmetric $m \times m$ matrix $A,$ let vech$[A]$ be the vector of its strictly lower triangular entries, which has length $q = m(m-1)/2.$ For any $\bu \in \mathbb{R}^{q},$ let vech$^{-1}[\bu]$ be the unique symmetric matrix whose strict lower-triangular entries are $\bu$ and whose diagonal entries are chosen so that every row sums to zero. In particular, if $\bu$ is nonpositive and bounded, then vech$^{-1}[\bu]\in\OL$, and each graph Laplacian $L$ is fully determined by vech$(L)$. 

In Scenario I, we consider an additive structure with 3-dimensional predictors, where we
generate the response variable
\begin{align}\label{scenario1-L}
    L = (1-\delta)\text{vech}^{-1}[-l_{1}, \cdots,-l_{q}] +\delta\text{vech}^{-1}[-\epsilon_{1}, \ldots, -\epsilon_{q}], \quad l_{j} = l_{j1} + l_{j2} + l_{j3}.
\end{align}
Here the $l_j$ are random variables, $0<\delta<1$ determines the contribution of additive noise $\epsilon_j$ leading to variability of $L$ and $\epsilon_{j} \sim \text{Bernoulli}(\epsilon)$ for  $j = 1,\ldots, q$.
Given the covariate vector $\bX$, we generate  the three additive components $l_{j1}, l_{j2}$ and $l_{j3}$ for $1 \le j \le q$ as
$\text{Beta}\left(\sin(\frac{\pi X_{1}}{2}), 1-\sin(\frac{\pi X_{1}}{2})\right)$, $\text{Beta}\left(X_{2}^{2}, 1-X_{2}^{2}\right)$ and $\text{Beta}\left(\exp(- X_{3}), 1-\exp(- X_{3})\right)$, respectively.
For the response variable $L$ generated in this way, all entries of the matrix function $m_{\OL}$
form an  additive function. Indeed,
\begin{align}\label{scenario1}
m_{\OL}(\bx)=(1-\delta)\text{vech}^{-1}[-\nu_1(\textbf{x}), \ldots, -\nu_1(\textbf{x})] + \delta \text{vech}^{-1}[-\epsilon, \ldots, -\epsilon],
\end{align}
where $\nu_1(\textbf{x}) = g_1(x_1) + g_2(x_2) +g_3(x_3)$ with $g_{1}(x_1) = \sin( \frac{\pi x_1}{2})$, $g_{2}(x_2) = x_2^2$, $g_3(x_3) = \exp(-x_3)$, so that  each entry of $m_{\OL}$ belongs to $\cH_+^\R$, implying  $m_{+,\OL}=m_{\OL}$. 

In Scenario II, we generate network samples from a weighted stochastic block model \citep{aicher2015learning}.
Consider a weighted stochastic block model with a vector of community membership $\bz = (\bz_1^\top, \bz_2^\top)^\top$, where $\bz_{i}$ is a vector of length $m_{i}$
with all elements equal to $i$, and $m_{1} + m_{2} = m$.
The presence of an edge between nodes within each block follows a Bernoulli distribution, parameterized by the corresponding entry in the matrix of edge probabilities
$\Theta = \big(\begin{smallmatrix}
  p_{11} & p_{12}\\
  p_{21} & p_{22}
\end{smallmatrix}\big)$ with $p_{12} = p_{21}.$ We consider the following additive regression model
\begin{align}\label{scenario2}
    m_{\OL}(\bx) &= (1-\delta)\text{vech}^{-1}\left[ -\text{vech}(\Lambda) \odot (\nu_2(\textbf{x}), \ldots, \nu_2(\textbf{x}))  \right] + \delta \text{vech}^{-1}[ - \epsilon, \ldots, -\epsilon],
\end{align}
where $\odot$ denotes the element-wise product,  $\Lambda = \big(\begin{smallmatrix}
  (p_{11})_{m_1 \times m_1} & (p_{12})_{m_1 \times m_2}\\
  (p_{21})_{m_2 \times m_1} & (p_{22})_{m_2 \times m_2}
\end{smallmatrix}\big)$,
$(p_{kl})_{m_k \times m_l}$ is an $m_k \times m_l$ matrix with all elements equal to $p_{kl},$ for $k,l = 1,2,$ and $\nu_2(\textbf{x}) = h_1(x_1) + h_2(x_2) +h_3(x_3)$ where $h_{1}(x_1) = \sin(\pi x_1)$, $h_{2}(x_2) = x_2^2$ and $h_3(x_3) = \exp(-x_3)$.
This model has a latent block structure.
We generate a response $L$ with the additive structure at \eqref{scenario2} as
\begin{align*}
    L = (1-\delta)\text{vech}^{-1}\left[ - \text{vech}(B) \odot (l_{1}, \ldots, l_{q}) \right]  + \delta\text{vech}^{-1}[-\epsilon_{1}, \ldots, -\epsilon_{q}],
\end{align*}
where $B = \big(\begin{smallmatrix}
  B_{11} & B_{12}\\
  B_{21} & B_{22}
\end{smallmatrix}\big),$ $B_{11}$ and $B_{22}$ are symmetric, $B_{12} = B^{\mathrm{T}}_{21},$ $B_{kl}$ is an $m_k \times m_l$ matrix with each element following Bernoulli($p_{kl}$) for $k, l = 1, 2,$ $\epsilon_{j}$ are the same as those in Scenario I, and $l_{j} = l_{j1} + l_{j2} + l_{j3}$.
The random components $l_{j1}, l_{j2}$ and $l_{j3}$ are generated from $\text{Beta}(\sin(\pi X_{1}), 1-\sin(\pi X_{1}))$, $\text{Beta}(X_{2}^{2}, 1-X_{2}^{2})$
and $\text{Beta}(\exp(- X_{3}), 1-\exp(- X_{3}))$. In this scenario, we also have $m_{+,\OL}=m_{\OL}$.

In Scenario III, we consider a regression setting similar to \eqref{scenario1} in Scenario I, but with a single-index structure. In this scenario, $\nu_1(\bx)$ in \eqref{scenario1}
is replaced by $\nu_3(\bx) = 1/2 + \sin(2\pi (\bx^{T}\bm{\theta_0})^2)/2$ where $\bm{\theta_0}$ is a $3$-dimensional unit vector with the first element $\theta_{01} > 0$.
The direction parameter $\bm{\theta_0}$ is drawn uniformly from the upper hemisphere $\mathbb{S}^{2}_{++} = \{(x, y, z) \in \R^3 \mid x>0,~ x^2 + y^2 + z^2 = 1 \}$.
The response $L$ is generated according to \eqref{scenario1-L} with $l_j$ now following
$\text{Beta}\left(\frac{1}{2} + \frac{1}{2} \sin(2\pi (\bx^{T}\bm{\theta_0})^2), \frac{1}{2} - \frac{1}{2} \sin(2\pi (\bx^{T}\bm{\theta_0})^2)\right)$ for
$j = 1,\ldots, q$.

Scenarios I and II generate random network responses for which the underlying Fr\'echet regression function is consistent with the  proposed additive \F regression model
at \eqref{def-m+Omega}.
Scenario III does not generate network responses aligned with an additive structure, as the nonlinear link function $(\bx^{T}\bm{\theta_0})^2$ introduces
interactions between the predictors. We include this scenario to evaluate the performance of the  additive \F regression model when the model assumptions are not satisfied.

Random samples of pairs $(\bX_{i}, L_i),$ $i = 1,\ldots, n$ are generated as $\textbf{X} = (X_{1},X_{2}, X_{3}) = (\Phi(V_{1}),\Phi(V_{2}),\Phi(V_{3}))$,
where $\Phi$ is the standard normal cumulative distribution function (CDF) and $\textbf{V} = (V_{1}, V_{2}, V_{3})^{T} \sim N_{3}(0, \Sigma)$
has a multivariate normal distribution with mean zero and a covariance matrix $\Sigma = (\sigma_{kl}),$ $\sigma_{kk} = 1,$ $k = 1,2,3$,
$\sigma_{12} = \sigma_{21} = \sigma_{23} = \sigma_{32} = 0.5,$ and $\sigma_{13} = \sigma_{31} = 0.3$.
The samples  $L_i$ are then generated as described for each scenario.
In all scenarios, $\epsilon = 0.5$ and $\delta=0$ or $\delta=0.05$, where for
Scenarios I and III,  the number of nodes is set as $m = 10$ and
for Scenario II, the block structure dimensions are chosen  as $m_1 = m_2 = 5$ and the edge probabilities as $p_{11} = p_{22} = 0.8$, $p_{12}=p_{21} = 0.6$.
For each setting, we generate $n = 100, 200, 300$ and $500$ samples $\{(\bX_{i},L_{i}): 1\leq i \leq n \}$.
We compare the performance of additive \F regression (AFR) with: 1) network regression (NR) based on graph Laplacians \citep{zhou2022network},
2)  single-index \F regression (SIF) \citep{bhattacharjee2023single},
and 3) manifold-valued regression (MR) proposed by \cite{severn2022manifold} and base these comparisons on  the integrated squared error (ISE)  using  the Frobenius metric $d_{F}$,
\begin{equation*}
    \text{ISE}_{b} = \int_{[0,1]^{3}} d_{F}^{2}({m_{\OL}(\bx)}, \hat{m}^{(b)}(\bx)) d\bx,
\end{equation*}
where $\hat{m}^{(b)}(\bx)$ denotes the regression estimate from the $b^{th}$ Monte Carlo (MC) sample. 

\begin{figure}[!ht]
    \centering
    \includegraphics[width=0.95\linewidth]{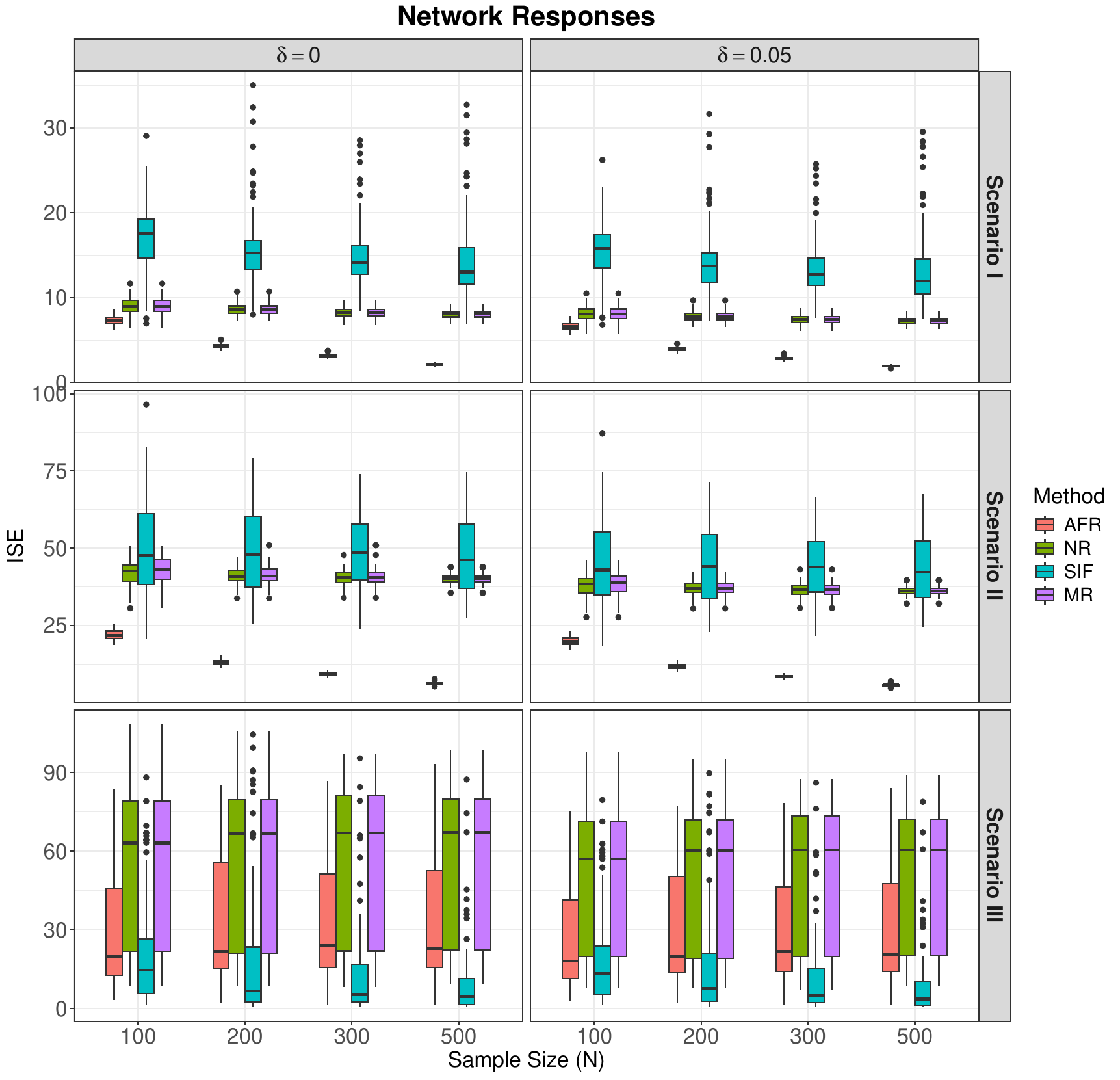}
    \caption{\small Boxplots of the ISE for $B = 200$ Monte Carlo simulations. Scenario I (Additive structure without block structure) and II (Additive structure with block structure),
    and Scenario III (Single-Index structure, not corresponding to additive structure).
    Four methods are compared: additive \F regression (AFR), network regression (NR), manifold-valued regression (MR) and single-index \F regression (SIF) for four sample sizes. } 
    \label{fig:net_sim}
\end{figure}
Figure \ref{fig:net_sim} illustrates the resulting boxplots for  ISE from $B = 200$ MC simulations across the three scenarios (I, II, and III).
In Scenarios I and II, where the underlying regression functions have additive forms, the proposed AFR model is seen to  substantially outperform the other three methods in terms of ISE, with decreasing  ISE  as the sample size $n$ increases.
In these scenarios, SIF exhibits relatively high variance because it first estimates a single-index direction and then applies local \F regression;
the lack of a well-defined direction in the true model leads to large variation in the ISE results across simulations.
In Scenario III, which is deliberately tailored to favor SIF and does not align with our model assumption,
SIF naturally performs best. Despite this, the proposed additive model still achieves the second-best median ISE among the four methods.

\subsection{Univariate distributional responses in the Wasserstein space}\label{sec:den_simul}

We consider here univariate distributional responses $Y$ situated in  the $2$-Wasserstein space  $(\Omega_{W}, d_{W})$ as described in Example 2 of Section \ref{sec:Add-F-reg-mean},
with Euclidean predictor vectors  $\bX \in \R^{p}$ and denote the  quantile function corresponding to a distribution $Z$ by $Q(Z)(\cdot)$. We
construct a Gaussian distributional response $Y$ as follows. Denoting by  $\Phi^{-1}(\cdot)$ the quantile function of the standard normal distribution, we set
\begin{equation*}
\E(Q(Y)(\cdot)|X = \bx) =  \mu_0(\bx) + \sigma_0(\bx) \Phi^{-1}(\cdot).
\end{equation*}
In this way, if $\mu_0$ and $\sigma_0$ are additive functions so that $\E(Q(Y)(s)|X = \cdot) \in \cH_+^\R$ for each $s \in [0,1]$,
then $m_{+,\Omega_W}(\bx)=m_{\Omega_W}(\bx)$ which is the distribution with $\E(Q(Y)(\cdot)|X = \bx)$ as its quantile function.
We generate $Q(Y)(\cdot)$ according to $Q(Y)(\cdot) = \mu(\bX) + \sigma(\bX) \Phi^{-1}(\cdot)$ with $\mu(\bx) = \mu_0(\bx) + \epsilon_\mu$,
$\sigma(\bx) = \sigma_0(\bx) + \epsilon_\sigma,$ and $\epsilon_\mu,\epsilon_\sigma  \sim TN_{[-1,1]}(0, 0.5^2)$, where $TN_{[a,b]}$ is a truncated normal distribution restricted to $[a, b]$. Note that the distributional response $Y$ that has $\mu(\bX) + \sigma(\bX) \Phi^{-1}(\cdot)$ as its quantile function is
the Gaussian distribution $N(\mu(\bX),\sigma^2(\bX))$.
Random samples of the covariates $\bX_i$ are generated in the same way as in \cref{sec:net_simul}.
We consider three \F regression scenarios:
\begin{itemize}
    \item Scenario I (Additive): $\mu_0(\bx) = x_1^2 + \sin(2\pi x_2) + \exp(-x_3)$ and $\sigma_0(\bx) = 3 +  x_1^2 + \sin(2\pi x_2) + \exp(-x_3)$.

    \item Scenario II (Linear): $\mu_0(\bx) =  2x_1 - x_2 + 3x_3$ and $\sigma_0(\bx) = 3 +  x_1 + x_2 - x_3.$

    \item Scenario III (Single Index): $\mu_0(\bx) = 1 +2\sin(2\pi(\bx^{T}\bm{\theta_0})^2)$ and $\sigma_0(\bx) = 2 + \exp(-\bx^{T}\bm{\theta_0})$,
    where $\bm{\theta_0}$ is a unit vector randomly drawn from the upper hemisphere $\mathbb{S}^{2}_{++}$ with the first element $\theta_{01} > 0$.
\end{itemize}

Scenarios I and II generate random response distributions such that the underlying \F regression function $m_{\Omega_W}$ at \eqref{Frechet-mean}
coincides with $m_{+, \Omega_W}$ at \eqref{def-m+Omega}.
In addition, Scenario II is specifically tailored to satisfy the assumptions of the global \F regression model  \citep{petersen2019frechet}.
Scenario III implements a  single‐index \F regression model  \citep{bhattacharjee2023single,ghosal2023frechet}, for which  $m_{+,\Omega_W} \neq  m_{\Omega_W}$ and the additive model assumption is not satisfied.
This scenario allows us to study the behavior of the proposed model when the assumptions are not satisfied.

In the second setting, we construct a non‐Gaussian distributional response $Y$. For this, we first generate distribution parameters $\mu$
and $\sigma$ according to Scenarios I–III,
and then apply a random transport map $T$ in the Wasserstein space. The transport map $T$ is chosen uniformly from the family
$\{T_{k}(a) = a - \sin(ka)/|k|: k \in \{\pm1, \pm 2\} \}$. Then the distributional response $Y$ is given by
$Y = T \# N(\mu(\bX),\sigma^2(\bX))$, where $\#$ denotes the push-forward operation.

Note that \cite{han2020additive} restrict additive distributional regression to responses supported on the compact domain, i.e., $[0,1],$ so that the transformed response space is bounded. However, the distributional space with the 2-Wasserstein metric $(\Omega_{W}, d_{W})$ may be totally bounded even for distributions with non-compact support. In the Gaussian case, every scenario we consider lies in $\left\{N(\mu, \sigma^2) \mid \mu \in \left[\mu_{\text{min}}, \mu_{\text{max}}\right],~ 0 < \sigma_{\text{min}} \leq \sigma \leq \sigma_{\text{max}} \right\},$ which can be covered by finitely many $d_{W}$-balls of any radius $\delta >0$ and is therefore totally bounded. For our non-Gaussian setting with $2$-Lipschitz transport $T,$ the space $\left\{T \# N(\mu, \sigma^2)\mid  \mu \in \left[\mu_{\text{min}}, \mu_{\text{max}}\right],~ 0 < \sigma_{\text{min}} \leq \sigma \leq \sigma_{\text{max}} \right\}$ can be covered by the same $d_{W}$-balls of radius $2\delta,$ thus is also totally bounded.

For each setting, we generate $n = 100, 200, 300$ and $500$ pairs of $(\bX,Y).$ We compare finite-sample performance with 1) the global \F regression (GFR) \citep{petersen2019frechet}, 2)
the single-index \F regression (SIF) \citep{bhattacharjee2023single} and 3) additive density regression (ADR) \citep{han2020additive} in terms of the ISE based on the Wasserstein distance $d_{W}$, defined by
\begin{equation*}
    \text{ISE}_{b} = \int_{[0,1]^{3}} d_{W}^{2}(m_{\Omega_W}(\bx), \hat{m}^{(b)}(\bx)) \, {\rm d}\bx.
\end{equation*}

\begin{figure}[!ht]
    \centering
    \includegraphics[width=0.95\linewidth]{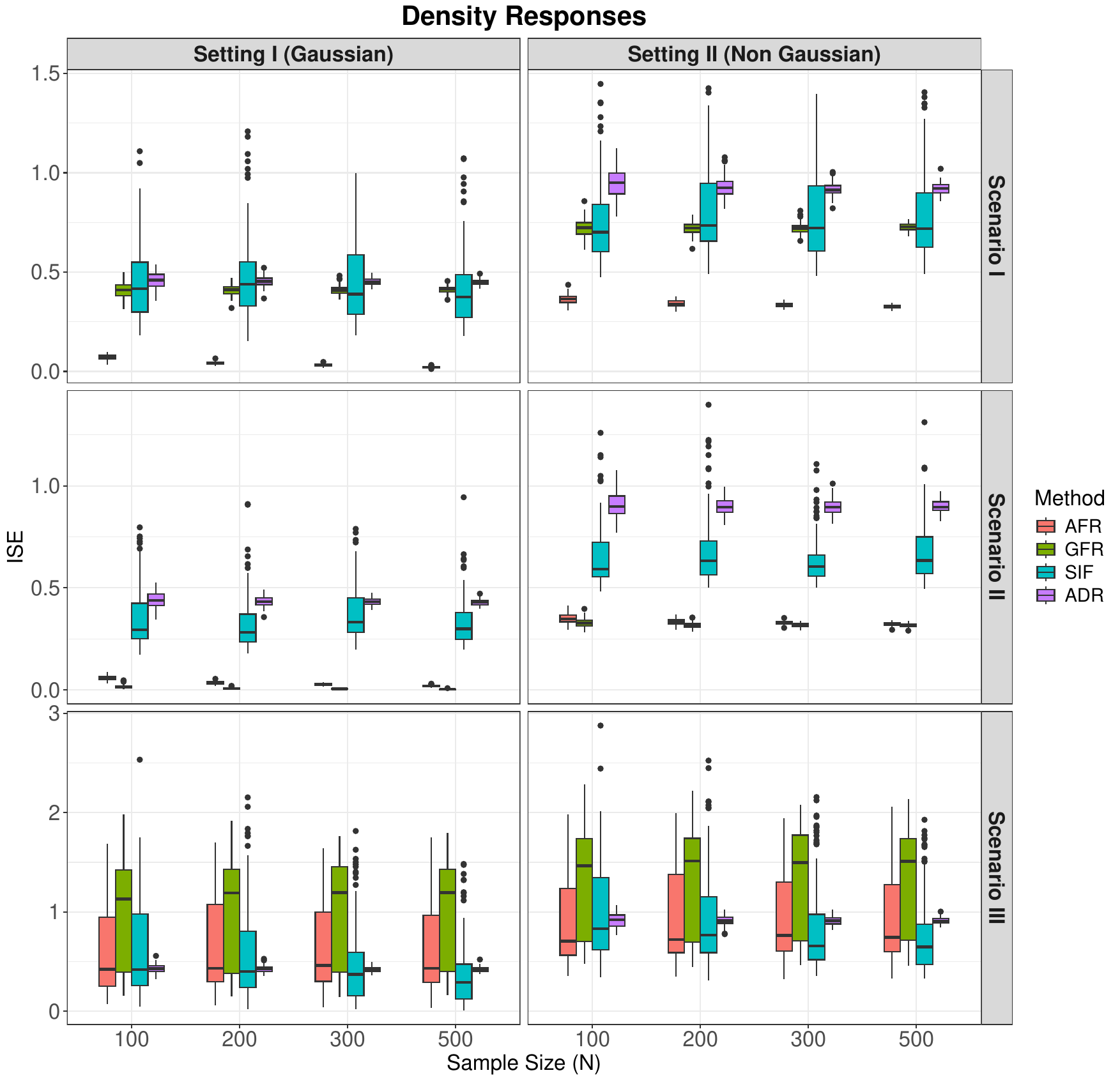}
    \caption{\small Boxplots of the ISE over $B = 200$ Monte Carlo simulations for Gaussian responses (left) and Non Gaussian responses (right). Four methods are compared: additive \F regression (AFR), global \F regression (GFR), single-index \F regression (SIF), and additive density regression (ADR) for  sample sizes $n = 100, 200, 300,$ and $500$.}
    \label{fig:den_sim}
\end{figure}

The left column of Figure \ref{fig:den_sim} presents boxplots of the ISE obtained from $B=200$ Monte Carlo runs, where the responses $Y$ follow Gaussian distributions under the three scenarios described above.
In Scenario I, where the underlying regression function has an additive form, the proposed additive approach produces significantly better results than the other methods.
Scenario II is specifically tailored for global Fr\'echet regression (GFR), so GFR naturally performs the best. But this model also satisfies the additive model assumptions and in this scenario, the proposed approach is seen to be
competitive compared to GFR and  demonstrates far stronger performance compared to  the other alternatives (SIF, ADR).

In both Scenarios I and II, ADR performs poorly, likely because it requires (i) an extrinsic additive form once a suitable transformation (e.g., an LQD or log-hazard transformation) is applied, and (ii) a compact domain (e.g., $[0, 1]$), neither of which holds in the present setting. Furthermore, SIF exhibits comparatively high variance because it first estimates a direction, then applies local Fr\'echet regression, and the absence of a well-defined direction in the true model leads to  large variation in the ISE across simulations.
Finally, in Scenario III, SIF would be expected to perform  best,  as this setting is tailored for SIF and
the underlying model does not align with AFR. Nonetheless,  the proposed AFR method  performs comparably to SIF
in terms of the median ISE.

The right column of Figure \ref{fig:den_sim} presents the results for the non-Gaussian setting, where the distributional response $Y$ is generated
via the random transport map $T$. Compared to the Gaussian results in the left column, the ISE values are generally higher across all scenarios. Nevertheless, the proposed model substantially outperforms the other approaches  in Scenario I, and its performance is comparable to the competing methods in Scenarios II and III.
Specifically, although Scenario II is based on the GFR structure, AFR and GFR exhibit similar ISE values.  AFR demonstrates even better performance than SIF in Scenario III for the smallest sample size $n = 100,$ as indicated by a lower median ISE.

\section{Data Applications}\label{sec:data_application}

\subsection{ADNI Brain Network}\label{sec:data_ADNI}

We used resting-state fMRI (rs-fMRI) data obtained from the Alzheimer's Disease Neuroimaging Initiative (ADNI) database (\url{https://adni.loni.usc.edu/}).
Network-based analyses of brain functional connectivity at the subject level typically rely on a specific spatial parcellation of the brain into a set of regions of interest (ROIs). The functional connectivity  between pairwise ROIs is usually quantified by Pearson correlation coefficients of the fMRI time series, leading to an $m \times m$ correlation matrix for  $m$ ROIs. Hard or soft thresholding is customarily applied to produce binary or weighted functional connectivity networks from the correlation matrices. In our analyses, we used density-based thresholding, where the threshold is allowed to vary from subject to subject to achieve a fixed connection density,  retaining only the strongest $15 \%$ of connections  \citep{zhou2022network}. The list of ROIs is in Table S.1 in Supplementary Material  S.5.

We analyzed blood-oxygen-level-dependent (BOLD) signals from the automated anatomical labeling (AAL) atlas to parcellate the whole brain into 90 ROIs, with 45 ROIs in each hemisphere. For the AAL atlas, we have $n = 347$ brain networks.
We consider two predictors ($p = 2$): The first predictor $X_1$ represents the ADAS-Cog-13 score (henceforth referred to as C score), which is a widely used measure of cognitive performance. A higher C score indicates worse cognitive function \citep{mohs1997development}. The second predictor $X_2$ denotes the level of phosphorylated tau (pTau) in cerebrospinal fluid (CSF), measured in picograms per milliliter (pg/mL), which  is a biomarker for Alzheimer's disease (AD). Its  association with AD pathology has been extensively investigated in recent studies \citep{karikari2022blood, rawat2022phosphorylated}. Participants in the study  had values for pTau ranging from 8.00 to 92.08 (pg/mL) and their C scores ranged from 1.00 to 65.67.

\begin{figure}[!ht]
    \centering
    \includegraphics[width=0.5\linewidth]{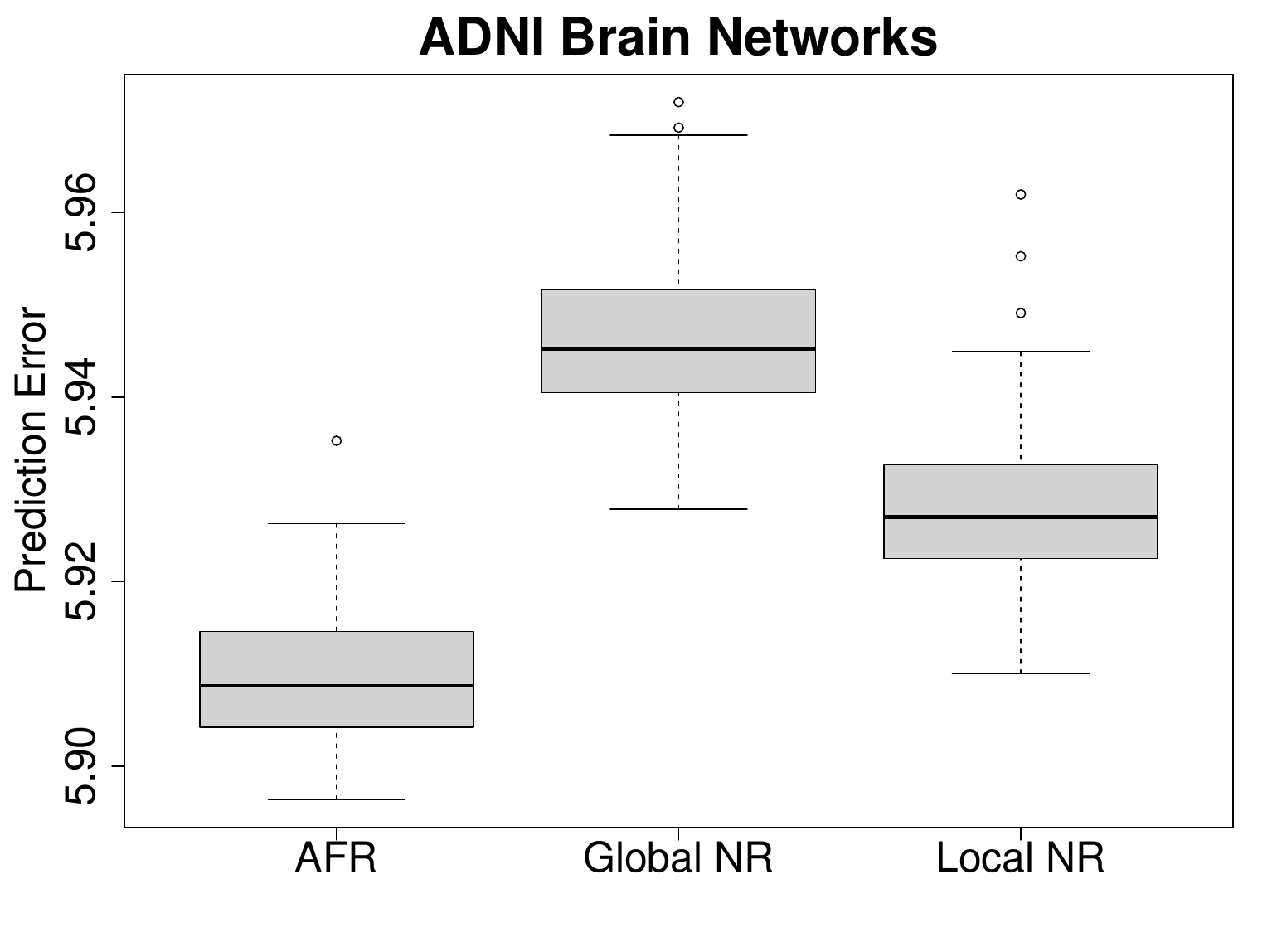}
    \caption{\small The 5-fold cross-validation prediction errors of ADNI brain network regression from $B=200$ random splits into 5 folds, comparing the proposed additive \F regression (AFR) with global and local network regression (NR).}
    \label{fig:adni_prederr}
\end{figure}

We compared the proposed additive \F regression approach (AFR) with the previously proposed global network regression (Global NR) and local network regression (Local NR) \citep{zhou2022network} using 5-fold cross-validation prediction error,
\begin{equation*}
    \text{Prediction Error} = \frac{1}{5}\sum_{k = 1}^{5} \text{CV}_{k}, \quad \text{CV}_{k} = \frac{1}{n_{k}}\sum_{i \in \text{Fold}_{k}} d^{2}_{F}(\hat{L}^{(-k)}_{i}, L_{i}),
\end{equation*}
where $\hat{L}^{(-k)}_{i}$ is the predicted value obtained by fitting a regression model trained on the four remaining folds with the $k$-th fold left out  and
evaluating the trained regression function at $\bX_i$ in the $k$-th fold.
Figure \ref{fig:adni_prederr} displays the resulting 5-fold cross-validation prediction errors obtained from 200 random splits into 5 folds, demonstrating that the proposed additive \F regression has the lowest  prediction error.

\begin{figure}[!ht]
    \centering
    \includegraphics[width=0.45\linewidth]{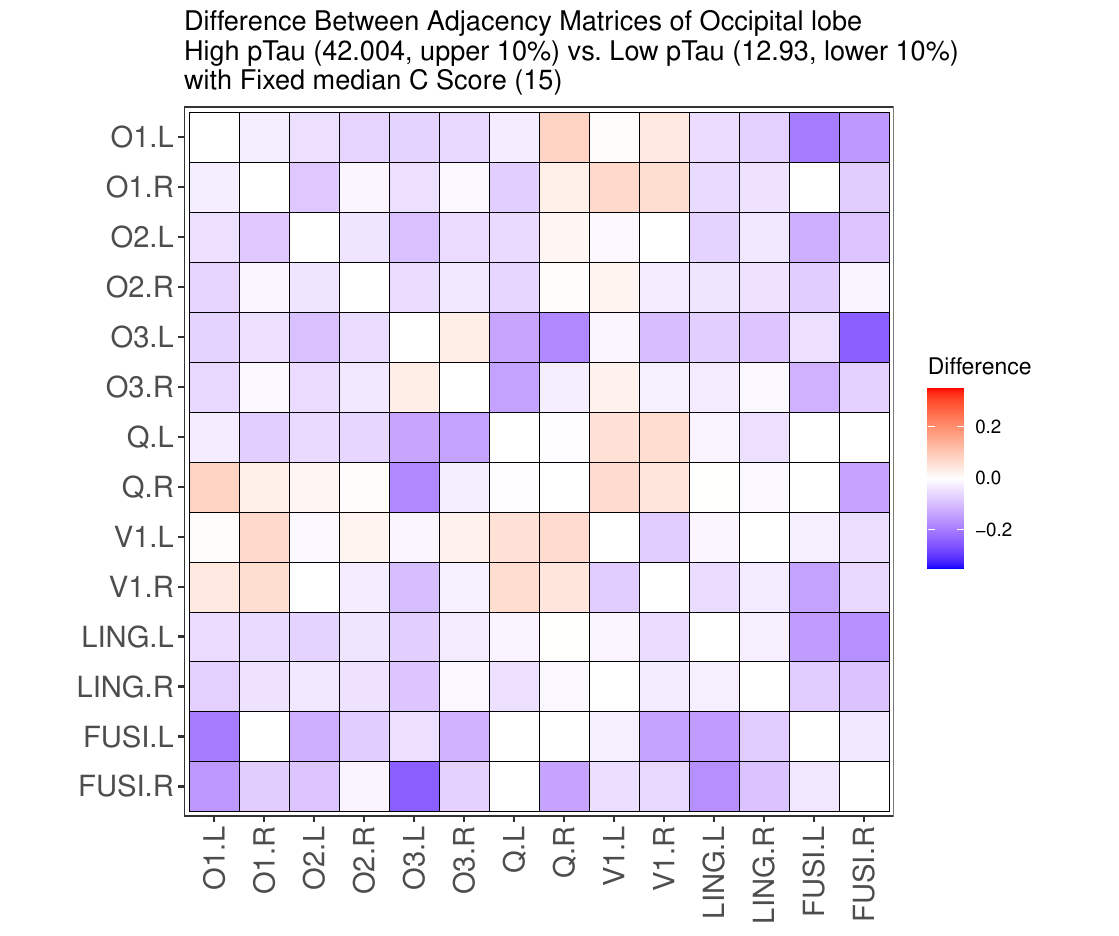}
    \includegraphics[width=0.45\linewidth]{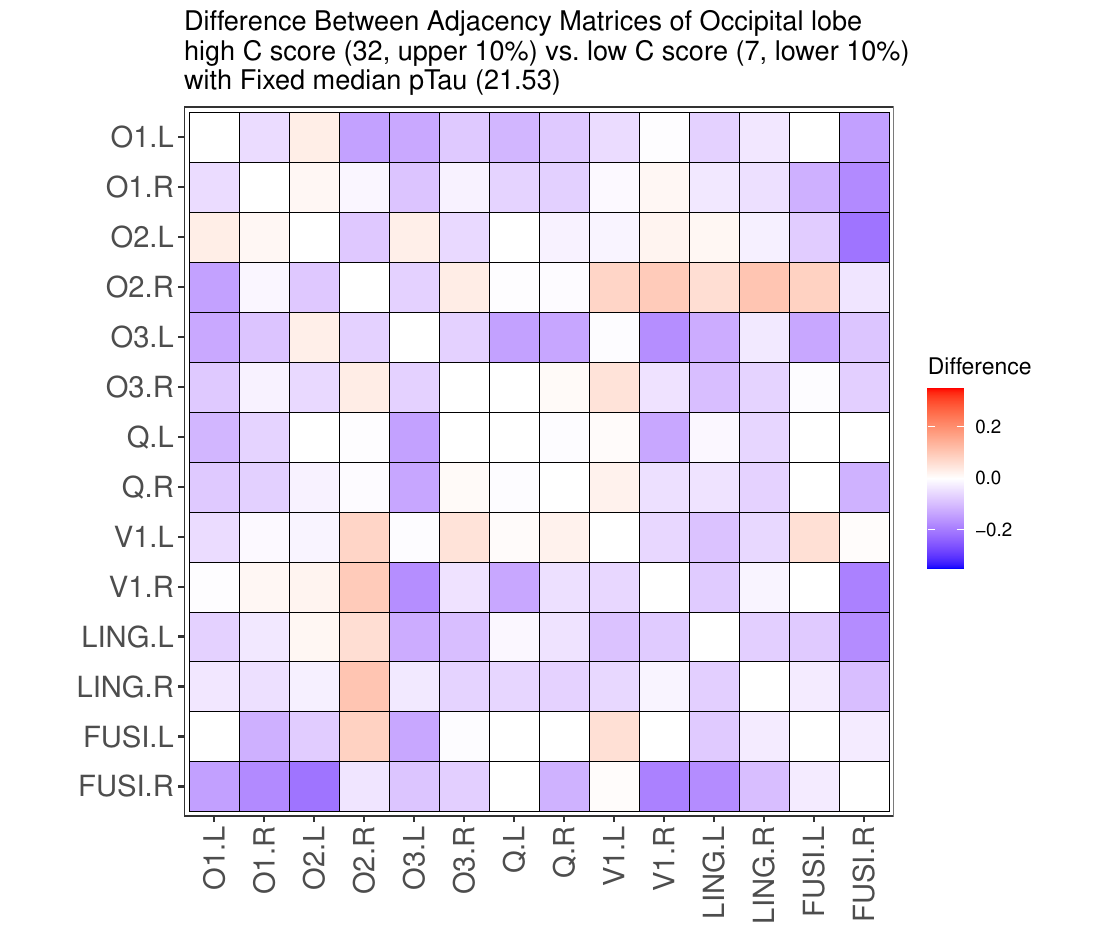}
    \caption{\small Difference between two predicted adjacency matrices for the occipital lobe with high pTau (42.0, upper 10\%) and low pTau (12.9, lower 10\%) for fixed median C score 15 (left)
    and with high C score (32, upper 10\%) and low C score (7, lower 10\%) for fixed median pTau 21.53 (right).}
    \label{fig:adni_occipital}
\end{figure}

To quantify the impact of covariates in the regression model on brain connectivity networks, we present entry-wise differences in the predicted adjacency matrices. Specifically, to examine the effect of the pTau variable, we fix the C score covariate at 15 (the median) and compare the additive \F regression estimates between high pTau (42.0, representing the upper 10\%) and low pTau (12.9, representing the lower 10\%) groups. Since we retained only the top $15 \%$ of the strongest connections, most elements in the adjacency matrix are 0. Given this sparsity of the adjacency matrix, the left panel of Figure \ref{fig:adni_occipital} reveals notable differences in the occipital lobe, where high pTau levels are associated with decreased connectivity. The right panel of Figure \ref{fig:adni_occipital} shows that the estimated network for the high C score group exhibits lower correlations among ROIs in the occipital lobe compared to the low C score group. A more detailed analysis  is provided in Supplementary Material  S.5.

\subsection{NHANES Data}\label{sec:data_NHANES}

The National Health and Nutrition Examination Survey (NHANES) is a large-scale, stratified, multistage survey carried out by the Centers for Disease Control. It provides a wide range of health and nutrition data for the U.S. population and is commonly used in statistical analyses of physical activity \citep{matabuena2023distributional}. Comprehensive information—including accelerometer records, as well as sociodemographic, clinical, and dietary data—is publicly available (\url{https://www.cdc.gov/nchs/nhanes/}).
Our analyses draw on data from NHANES cohorts spanning 2011 to 2014.

In this study, each participant had at least 10 hours per day of accelerometer monitoring for a minimum of four days. For the $i$-th participant, raw accelerometer data take the form of time–activity pairs $(t_{ij}, l_{ij}),$ where $j = 1, \ldots, n_{i}.$ To convert these functional measurements into distributional data, we represent each participant's activity distribution $Y_i$ by an empirical quantile function $Q(Y_i)(\cdot) = \inf\{u \in \mathbb{R}\mid \hat{F}_i(u) \geq \cdot \},$ with $\hat{F}_i(u) = \frac{1}{n_{i}} \sum_{j=1}^{n_i} \textbf{I}(l_{ij} \leq u).$ The resulting response space is then the set of all such distributions, which we equip with the Wasserstein metric $d_{W}$.
Related approaches for the analysis of accelerometer data include   \cite{chang2022empirical, lin2023causal, ghosal2023frechet,song2024multi}.

We included three covariates in our analysis: (i) Age, (ii) Body Mass Index (BMI), and (iii) Healthy Eating Index (HEI).
Participants were between 65 and 80 years old at the time of measurement. The BMI, measured in ${\rm kg/m^2}$, ranged from 18.5 to 40,
encompassing categories from healthy weight to obesity.
The HEI, a continuous score ranging from 20.7 to 94.4, reflects overall diet quality; higher scores indicate better dietary habits.
A total of 1012 individuals were included in this analysis.
We compared the proposed additive \F regression with global \F regression \citep{petersen2019frechet}, assessing  predictive performance through a 5-fold cross-validation procedure with  cross-validation prediction error
\begin{equation*}
    \text{PredErr} = \frac{1}{5}\sum_{k = 1}^{5} \text{CV}_{k}, \quad \text{CV}_{k} = \frac{1}{n_{k}}\sum_{i \in \text{Fold}_{k}} d^{2}_{W}(\hat{Y}^{(-k)}_{i}, Y_{i}),
\end{equation*}
where $\hat{Y}^{(-k)}_{i}$ is defined analogously to $\hat{L}^{(-k)}_{i}$ in \cref{sec:data_ADNI}.
Table \ref{tbl:nhanes} presents  the averages of   PredErr and their standard errors.

\begin{table*}[h!]
\caption{Average values of PredErr with their standard errors in parentheses based on $200$ random cross-validation splits, comparing the proposed additive \F regression (AFR)
and global \F regression (GFR)  \citep{petersen2019frechet} for the NHANES data in the Wasserstein space.}
\label{tbl:nhanes}
\begin{tabular}{lcc}
\toprule
Method & AFR & GFR \\
\midrule
Average PredErr (SE) & {\bf 10.342 (0.002)}\ & 14.587 (0.002) \\
\bottomrule
\end{tabular}
\end{table*}

Let $\bx = (x_{\text{Age}}, x_{\text{BMI}}, x_{\text{HEI}})$ denote the covariate vector and  $F(m_{+,\Omega}(\bx))(\cdot)$ the CDF representation of the predicted distributional response ${m}_{+,\Omega}(\bx)$ in \eqref{def-m+Omega}.
To visualize the predicted physical activity distributions, we employ marginal exceedance functions \citep{kundu2025exceedance}. These  focus on the effect of a single covariate while holding the other covariates fixed and are defined as 
\begin{equation}
\hat{S}(\tau, x_{\text{target}}) = 1 - F(\hatm_{+,\Omega}(x_{\text{target}}, \bx_{0, \text{others}}))(\tau),
\label{exceed_ftn}
\end{equation}
quantifying the probability that the predicted activity level exceeds a specified threshold $\tau$ for a given value of the target covariate.
For instance, to investigate the effect of Age, we set $x_{\text{target}} = x_{\text{Age}}$ and fix the remaining covariates as $\bx_{0, \text{others}} = (x_{0,\text{BMI}}, x_{0,\text{HEI}})$.
Similarly, marginal exceedance functions $\hat{S}(\tau, x_{\text{BMI}})$ and $\hat{S}(\tau, x_{\text{HEI}})$ can be defined by fixing the other two covariates and varying BMI or HEI, respectively.

\begin{figure}[h!]
    \centering
    \includegraphics[width=0.975\linewidth]{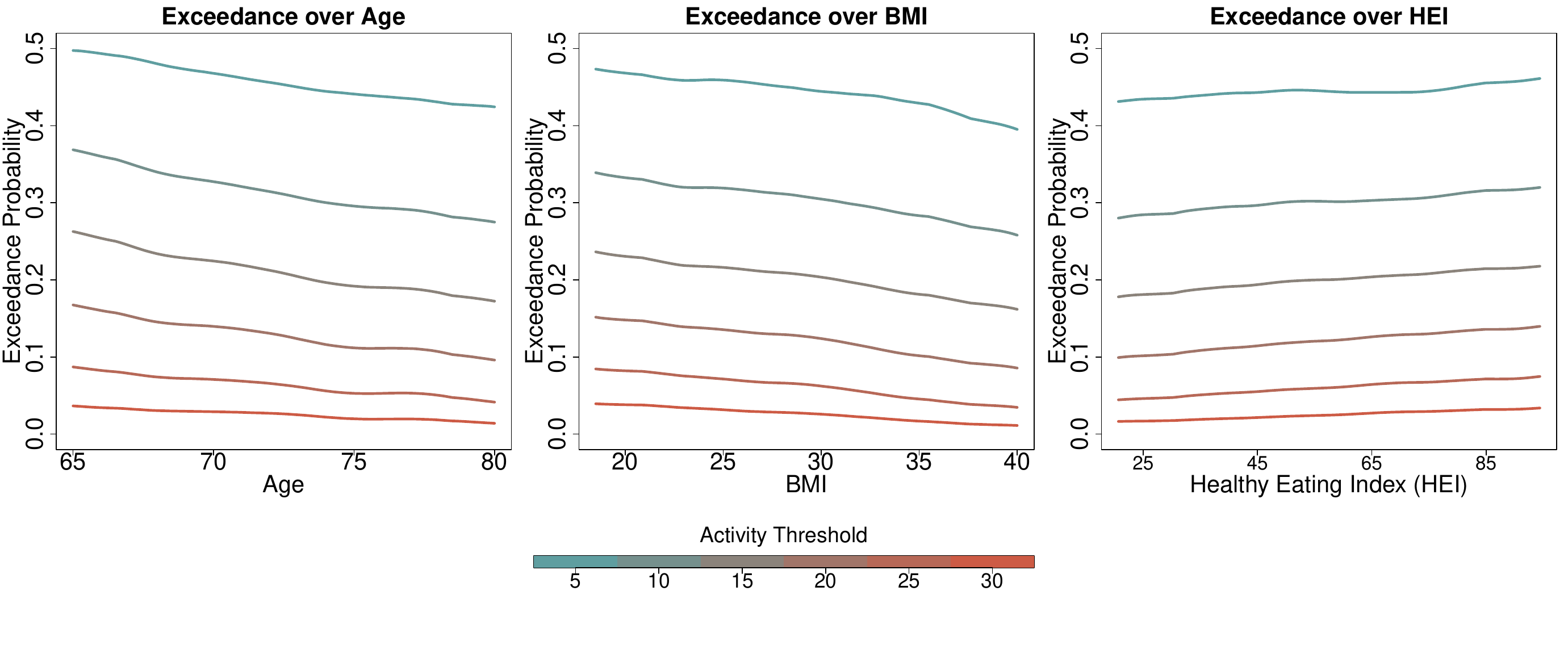}
    \caption{\small Predicted physical activity distributions derived from the proposed additive Fr\'echet regression model $\hat{m}_{+,\Omega}$ in \eqref{def-m+Omega-sample}, visualized as marginal exceedance functions $\hat{S}(\tau, x)$ in \eqref{exceed_ftn} for Age (left), BMI (middle), and HEI (right). Each curve corresponds to an activity threshold $\tau \in \{5, 10, 15, 20, 25,30\}$, with color ranging from blue (low thresholds) to red (high thresholds). For marginal exceedance representation of each covariate, the remaining predictors are held fixed at their median values.}
    \label{fig:nhanes_predicted}
\end{figure}

To examine the individual effects of each predictor using marginal exceedance functions, we fix the remaining covariates at their median values: $\text{Age} = 72, \text{BMI} = 27.6$, and $\text{HEI} = 58.3$.
Figure \ref{fig:nhanes_predicted} presents the marginal exceedance functions $\hat{S}(\tau, x)$ in \eqref{exceed_ftn} for Age (left), BMI (middle), and HEI (right), using activity thresholds $\tau = 5, 10, 15, 20, 25$, and $30$.
These exceedance functions are derived from the predicted additive \F regression $\hat{m}_{+,\Omega}$ in \eqref{def-m+Omega-sample}.
For the Age covariate, the marginal exceedance functions consistently shift downward as age increases, across all threshold levels $\tau$, indicating a decline in physical activity with advancing age.
For BMI, the functions also exhibit a general downward trend, with a notably steeper decline observed around $\text{BMI} = 35$.
This suggests that individuals with BMI above $35$  experience a more rapid decrease in physical activity compared to those with lower BMI.
In contrast, the marginal exceedance function for HEI shows an upward trend, implying that individuals with higher HEI scores tend to have greater physical activity levels.
This may be because individuals with healthier dietary habits tend to be more physically active.

The NHANES data were reanalyzed in Section S.8 in the Supplementary Material,
where we use the Fisher-Rao metric \citep{baue:16,dai:22} instead of the 2-Wasserstein metric to examine the sensitivity of the results to the choice of the metric $d$. The overall patterns obtained under the Fisher-Rao metric are found to remain similar to those obtained under the 2-Wasserstein metric.

\section{Discussion}\label{sec:discussion}

We develop an intrinsic additive Fr\'echet regression technique to address the curse of dimensionality when modeling a random object response in the presence of  multivariate continuous Euclidean covariates,
by viewing the weight function and its role in additive regression through a new lens.
The resulting  additive Fr\'echet regression estimator achieves univariate rates of convergence regardless of the fixed dimension $p$ of the covariate, thanks to
the intrinsic additive structure. This convergence holds both locally, i.e., pointwise for each fixed value of the covariate,  and also uniformly  across the  domain of the covariate vector.
Establishing these results is far more challenging than the existing theory of additive modeling for Euclidean or Hilbert-space-valued responses and requires analyzing various empirical processes indexed by function classes
arising from repeated applications of projection operators. This is a new feature that has been absent in the existing theory of smooth backfitting.

Classical additive regression with Euclidean responses has primarily been studied in settings where the predictor dimension $p$ is fixed. Of interest for future research is the case where $p$ 
diverges. 
Recent work by \citet{lee2024efficient} proposes an estimation procedure that combines smooth backfitting with a functional Lasso penalty to address 
a setting in which $p$ diverges, for the much simpler scalar case. Its further extension to the case with random object responses will be challenging.
In particular, Assumption (U1) imposes a uniform identifiability condition over $\bx \in [0,1]^p$ with a strictly positive lower bound $c(\epsilon).$  As $p$ gets larger but is still fixed,  this bound may deteriorate; while  univariate rates of convergence are preserved,  the constants might get larger. This will need to be addressed when one considers 
a diverging predictor dimension $p$. A rigorous treatment of these issues lies beyond the scope of the current paper.
Another recent work by \citet{moon2026ident} introduces a new formulation of additive interaction regression that enables the correct interpretation of interaction effects in nonparametric regression for the scalar response setting. Extending this 
to the case random object responses is an open problem of considerable interest.

Another promising future research direction is the study of a   combination of both linear and additive structure in the Fr\'echet regression function as a starting point  towards a more general semiparametric Fr\'echet regression framework, combining   the ideas of the additive Fr\'echet regression of the present work with the global Fr\'echet regression approach of \cite{petersen2019frechet}.
Such a generalized approach would  be particularly useful for modeling regression relationships between a response variable residing in a metric space
and covariates that include both continuous and binary or discrete variables.
Another challenge for future research is to incorporate an additive structure for non-Euclidean covariates residing in general metric spaces; also compare  \cite{jeon2021additive}.

%
%


\section*{Funding}
The research of Hans-Georg M\"uller was supported in part by NSF Grant DMS-2310450.
Changwon Choi’s research was supported by Basic Science Research Program through the National Research Foundation of Korea (NRF) funded by the Ministry of Education (No. RS-2024-00446677).
Byeong U. Park’s research was supported by the National Research Foundation (NRF) grant funded by Korea government (MSIT) (No. RS-2024-00338150, RS-2026-25606859).

\section*{Supplementary Material}
The Supplementary Material contains proofs of Theorems \ref{consistency}--\ref{uniform var property}, additional lemmas along with their proofs, further details on the ADNI brain network data analysis and the bandwidth selection algorithm of the additive \F regression model. It also features graphical illustrations of the additive weights that form the core of the proposed additive \F regression for the case $p=2$ of two-dimensional predictors and a sensitivity
analysis to gauge the effects of the choice of the metric $d$.

\section*{Acknowledgments} We wish to thank the Editor, the Associate Editor and the referees for helpful comments that led to an improved version of the paper.  Data used in preparation of this article were obtained from the Alzheimer’s Disease Neuroimaging Initiative (ADNI) database. As such, the investigators within the ADNI contributed to the design and implementation of ADNI and/or provided data but did not participate in analysis or writing of this report. A complete listing of ADNI investigators and databases can be found at \url{http://adni.loni.usc.edu}.


\bibliographystyle{imsart-nameyear} 
\bibliography{mybibfile}       

\begin{thebibliography}{50}

\bibitem[\protect\citeauthoryear{Afsari}{2011}]{afsari2011riemannian}
\begin{barticle}[author]
\bauthor{\bsnm{Afsari},~\bfnm{Bijan}\binits{B.}}
(\byear{2011}).
\btitle{Riemannian ${L}^{p}$ center of mass: existence, uniqueness, and
  convexity}.
\bjournal{Proceedings of the American Mathematical Society}
\bvolume{139}
\bpages{655--673}.
\end{barticle}
\endbibitem

\bibitem[\protect\citeauthoryear{Aicher, Jacobs and
  Clauset}{2015}]{aicher2015learning}
\begin{barticle}[author]
\bauthor{\bsnm{Aicher},~\bfnm{Christopher}\binits{C.}},
  \bauthor{\bsnm{Jacobs},~\bfnm{Abigail~Z}\binits{A.~Z.}} \AND
  \bauthor{\bsnm{Clauset},~\bfnm{Aaron}\binits{A.}}
(\byear{2015}).
\btitle{Learning latent block structure in weighted networks}.
\bjournal{Journal of Complex Networks}
\bvolume{3}
\bpages{221--248}.
\end{barticle}
\endbibitem

\bibitem[\protect\citeauthoryear{Bauer, Bruveris and Michor}{2016}]{baue:16}
\begin{barticle}[author]
\bauthor{\bsnm{Bauer},~\bfnm{Martin}\binits{M.}},
  \bauthor{\bsnm{Bruveris},~\bfnm{Martins}\binits{M.}} \AND
  \bauthor{\bsnm{Michor},~\bfnm{Peter~W}\binits{P.~W.}}
(\byear{2016}).
\btitle{{Uniqueness of the {Fisher--Rao} metric on the space of smooth
  densities}}.
\bjournal{Bulletin of the London Mathematical Society}
\bvolume{48}
\bpages{499--506}.
\end{barticle}
\endbibitem

\bibitem[\protect\citeauthoryear{Bhattacharjee, Li and
  Xue}{2025}]{bhattacharjee2025nonlinear}
\begin{barticle}[author]
\bauthor{\bsnm{Bhattacharjee},~\bfnm{Satarupa}\binits{S.}},
  \bauthor{\bsnm{Li},~\bfnm{Bing}\binits{B.}} \AND
  \bauthor{\bsnm{Xue},~\bfnm{Lingzhou}\binits{L.}}
(\byear{2025}).
\btitle{Nonlinear global Fr{\'e}chet regression for random objects via weak
  conditional expectation}.
\bjournal{The Annals of Statistics}
\bvolume{53}
\bpages{117--143}.
\end{barticle}
\endbibitem

\bibitem[\protect\citeauthoryear{Bhattacharjee and
  M{\"u}ller}{2023}]{bhattacharjee2023single}
\begin{barticle}[author]
\bauthor{\bsnm{Bhattacharjee},~\bfnm{Satarupa}\binits{S.}} \AND
  \bauthor{\bsnm{M{\"u}ller},~\bfnm{Hans-Georg}\binits{H.-G.}}
(\byear{2023}).
\btitle{Single index {F}r{\'e}chet regression}.
\bjournal{The Annals of Statistics}
\bvolume{51}
\bpages{1770--1798}.
\end{barticle}
\endbibitem

\bibitem[\protect\citeauthoryear{Bissantz et~al.}{2016}]{bissantz2016smooth}
\begin{barticle}[author]
\bauthor{\bsnm{Bissantz},~\bfnm{Nicolai}\binits{N.}},
  \bauthor{\bsnm{Dette},~\bfnm{Holger}\binits{H.}},
  \bauthor{\bsnm{Hildebrandt},~\bfnm{Thimo}\binits{T.}} \AND
  \bauthor{\bsnm{Bissantz},~\bfnm{Kathrin}\binits{K.}}
(\byear{2016}).
\btitle{Smooth backfitting in additive inverse regression}.
\bjournal{Annals of the Institute of Statistical Mathematics}
\bvolume{68}
\bpages{827--853}.
\end{barticle}
\endbibitem

\bibitem[\protect\citeauthoryear{Buja, Hastie and
  Tibshirani}{1989}]{buja1989linear}
\begin{barticle}[author]
\bauthor{\bsnm{Buja},~\bfnm{Andreas}\binits{A.}},
  \bauthor{\bsnm{Hastie},~\bfnm{Trevor}\binits{T.}} \AND
  \bauthor{\bsnm{Tibshirani},~\bfnm{Robert}\binits{R.}}
(\byear{1989}).
\btitle{Linear smoothers and additive models}.
\bjournal{The Annals of Statistics}
\bvolume{17}
\bpages{453--510}.
\end{barticle}
\endbibitem

\bibitem[\protect\citeauthoryear{Chang and McKeague}{2022}]{chang2022empirical}
\begin{barticle}[author]
\bauthor{\bsnm{Chang},~\bfnm{Hsin-wen}\binits{H.-w.}} \AND
  \bauthor{\bsnm{McKeague},~\bfnm{Ian~W}\binits{I.~W.}}
(\byear{2022}).
\btitle{Empirical likelihood-based inference for functional means with
  application to wearable device data}.
\bjournal{Journal of the Royal Statistical Society Series B: Statistical
  Methodology}
\bvolume{84}
\bpages{1947--1968}.
\end{barticle}
\endbibitem

\bibitem[\protect\citeauthoryear{Chen, Lin and M{\"u}ller}{2023}]{mull:23:2}
\begin{barticle}[author]
\bauthor{\bsnm{Chen},~\bfnm{Yaqing}\binits{Y.}},
  \bauthor{\bsnm{Lin},~\bfnm{Zhenhua}\binits{Z.}} \AND
  \bauthor{\bsnm{M{\"u}ller},~\bfnm{Hans-Georg}\binits{H.-G.}}
(\byear{2023}).
\btitle{Wasserstein regression}.
\bjournal{Journal of the American Statistical Association}
\bvolume{118}
\bpages{869--882}.
\end{barticle}
\endbibitem

\bibitem[\protect\citeauthoryear{Chen and M{\"u}ller}{2022}]{chen2022uniform}
\begin{barticle}[author]
\bauthor{\bsnm{Chen},~\bfnm{Yaqing}\binits{Y.}} \AND
  \bauthor{\bsnm{M{\"u}ller},~\bfnm{Hans-Georg}\binits{H.-G.}}
(\byear{2022}).
\btitle{Uniform convergence of local Fr{\'e}chet regression with applications
  to locating extrema and time warping for metric space valued trajectories}.
\bjournal{The Annals of Statistics}
\bvolume{50}
\bpages{1573--1592}.
\end{barticle}
\endbibitem

\bibitem[\protect\citeauthoryear{Cho et~al.}{2024}]{cho2024partially}
\begin{barticle}[author]
\bauthor{\bsnm{Cho},~\bfnm{Sungho}\binits{S.}},
  \bauthor{\bsnm{Jeon},~\bfnm{Jeong~Min}\binits{J.~M.}},
  \bauthor{\bsnm{Kim},~\bfnm{Dongwoo}\binits{D.}},
  \bauthor{\bsnm{Yu},~\bfnm{Kyusang}\binits{K.}} \AND
  \bauthor{\bsnm{Park},~\bfnm{Byeong~U}\binits{B.~U.}}
(\byear{2024}).
\btitle{Partially linear additive regression with a general {H}ilbertian
  response}.
\bjournal{Journal of the American Statistical Association}
\bvolume{119}
\bpages{942--956}.
\end{barticle}
\endbibitem

\bibitem[\protect\citeauthoryear{Choi et~al.}{2026}]{choi2026AFR}
\begin{barticle}[author]
\bauthor{\bsnm{Choi},~\bfnm{Changwon}\binits{C.}},
  \bauthor{\bsnm{Song},~\bfnm{Wookyeong}\binits{W.}},
  \bauthor{\bsnm{M{\"u}ller},~\bfnm{Hans-Georg}\binits{H.-G.}} \AND
  \bauthor{\bsnm{Park},~\bfnm{Byeong~U.}\binits{B.~U.}}
(\byear{2026}).
\btitle{Supplement to ``Additive Fr\'echet regression for random objects''}.
\end{barticle}
\endbibitem

\bibitem[\protect\citeauthoryear{Dai}{2022}]{dai:22}
\begin{barticle}[author]
\bauthor{\bsnm{Dai},~\bfnm{Xiongtao}\binits{X.}}
(\byear{2022}).
\btitle{Statistical inference on the {H}ilbert sphere with application to
  random densities}.
\bjournal{Electronic Journal of Statistics}
\bvolume{16}
\bpages{700-736}.
\end{barticle}
\endbibitem

\bibitem[\protect\citeauthoryear{Dubey and
  M{\"u}ller}{2022}]{dubey2022modeling}
\begin{barticle}[author]
\bauthor{\bsnm{Dubey},~\bfnm{Paromita}\binits{P.}} \AND
  \bauthor{\bsnm{M{\"u}ller},~\bfnm{Hans-Georg}\binits{H.-G.}}
(\byear{2022}).
\btitle{Modeling time-varying random objects and dynamic networks}.
\bjournal{Journal of the American Statistical Association}
\bvolume{117}
\bpages{2252--2267}.
\end{barticle}
\endbibitem

\bibitem[\protect\citeauthoryear{Fornito, Zalesky and
  Bullmore}{2016}]{fornito2016fundamentals}
\begin{bbook}[author]
\bauthor{\bsnm{Fornito},~\bfnm{Alex}\binits{A.}},
  \bauthor{\bsnm{Zalesky},~\bfnm{Andrew}\binits{A.}} \AND
  \bauthor{\bsnm{Bullmore},~\bfnm{Edward}\binits{E.}}
(\byear{2016}).
\btitle{Fundamentals of {B}rain {N}etwork {A}nalysis}.
\bpublisher{Academic press}.
\end{bbook}
\endbibitem

\bibitem[\protect\citeauthoryear{Ghosal, Meiring and
  Petersen}{2023}]{ghosal2023frechet}
\begin{barticle}[author]
\bauthor{\bsnm{Ghosal},~\bfnm{Aritra}\binits{A.}},
  \bauthor{\bsnm{Meiring},~\bfnm{Wendy}\binits{W.}} \AND
  \bauthor{\bsnm{Petersen},~\bfnm{Alexander}\binits{A.}}
(\byear{2023}).
\btitle{Fr{\'e}chet single index models for object response regression}.
\bjournal{Electronic Journal of Statistics}
\bvolume{17}
\bpages{1074--1112}.
\end{barticle}
\endbibitem

\bibitem[\protect\citeauthoryear{Groisser}{2004}]{groisser2004newton}
\begin{barticle}[author]
\bauthor{\bsnm{Groisser},~\bfnm{David}\binits{D.}}
(\byear{2004}).
\btitle{Newton's method, zeroes of vector fields, and the {R}iemannian center
  of mass}.
\bjournal{Adv. in Appl. Math.}
\bvolume{33}
\bpages{95--135}.
\bdoi{10.1016/j.aam.2003.08.003}
\bmrnumber{2064359}
\end{barticle}
\endbibitem

\bibitem[\protect\citeauthoryear{Han, M{\"u}ller and
  Park}{2020}]{han2020additive}
\begin{barticle}[author]
\bauthor{\bsnm{Han},~\bfnm{Kyunghee}\binits{K.}},
  \bauthor{\bsnm{M{\"u}ller},~\bfnm{Hans-Georg}\binits{H.-G.}} \AND
  \bauthor{\bsnm{Park},~\bfnm{Byeong~U}\binits{B.~U.}}
(\byear{2020}).
\btitle{Additive functional regression for densities as responses}.
\bjournal{Journal of the American Statistical Association}
\bvolume{115}
\bpages{997--1010}.
\end{barticle}
\endbibitem

\bibitem[\protect\citeauthoryear{Hong et~al.}{2025}]{hong2025robust}
\begin{barticle}[author]
\bauthor{\bsnm{Hong},~\bfnm{Houren}\binits{H.}},
  \bauthor{\bsnm{Scealy},~\bfnm{Janice~L}\binits{J.~L.}},
  \bauthor{\bsnm{Wood},~\bfnm{Andrew~TA}\binits{A.~T.}} \AND
  \bauthor{\bsnm{Yang},~\bfnm{Yanrong}\binits{Y.}}
(\byear{2025}).
\btitle{A robust extrinsic single-index model for spherical data}.
\bjournal{arXiv preprint arXiv:2503.24003}.
\end{barticle}
\endbibitem

\bibitem[\protect\citeauthoryear{Jeon and Park}{2020}]{jeon2020additive}
\begin{barticle}[author]
\bauthor{\bsnm{Jeon},~\bfnm{JEONG~MIN}\binits{J.~M.}} \AND
  \bauthor{\bsnm{Park},~\bfnm{BYEONG~U}\binits{B.~U.}}
(\byear{2020}).
\btitle{Additive regression with {H}ilbertian responses}.
\bjournal{The Annals of Statistics}
\bvolume{48}
\bpages{2671--2697}.
\end{barticle}
\endbibitem

\bibitem[\protect\citeauthoryear{Jeon, Park and
  Van~Keilegom}{2021}]{jeon2021additive}
\begin{barticle}[author]
\bauthor{\bsnm{Jeon},~\bfnm{Jeong~Min}\binits{J.~M.}},
  \bauthor{\bsnm{Park},~\bfnm{Byeong~U.}\binits{B.~U.}} \AND
  \bauthor{\bsnm{Van~Keilegom},~\bfnm{Ingrid}\binits{I.}}
(\byear{2021}).
\btitle{Additive regression for non-{E}uclidean responses and predictors}.
\bjournal{Ann. Statist.}
\bvolume{49}
\bpages{2611--2641}.
\bdoi{10.1214/21-aos2048}
\bmrnumber{4338377}
\end{barticle}
\endbibitem

\bibitem[\protect\citeauthoryear{Karikari et~al.}{2022}]{karikari2022blood}
\begin{barticle}[author]
\bauthor{\bsnm{Karikari},~\bfnm{Thomas~K}\binits{T.~K.}},
  \bauthor{\bsnm{Ashton},~\bfnm{Nicholas~J}\binits{N.~J.}},
  \bauthor{\bsnm{Brinkmalm},~\bfnm{Gunnar}\binits{G.}},
  \bauthor{\bsnm{Brum},~\bfnm{Wagner~S}\binits{W.~S.}},
  \bauthor{\bsnm{Benedet},~\bfnm{Andr{\'e}a~L}\binits{A.~L.}},
  \bauthor{\bsnm{Montoliu-Gaya},~\bfnm{Laia}\binits{L.}},
  \bauthor{\bsnm{Lantero-Rodriguez},~\bfnm{Juan}\binits{J.}},
  \bauthor{\bsnm{Pascoal},~\bfnm{Tharick~Ali}\binits{T.~A.}},
  \bauthor{\bsnm{Su{\'a}rez-Calvet},~\bfnm{Marc}\binits{M.}},
  \bauthor{\bsnm{Rosa-Neto},~\bfnm{Pedro}\binits{P.}} \betal{et~al.}
(\byear{2022}).
\btitle{Blood phospho-tau in Alzheimer disease: analysis, interpretation, and
  clinical utility}.
\bjournal{Nature Reviews Neurology}
\bvolume{18}
\bpages{400--418}.
\end{barticle}
\endbibitem

\bibitem[\protect\citeauthoryear{Kundu, Zhou and
  M{\"u}ller}{2025}]{kundu2025exceedance}
\begin{barticle}[author]
\bauthor{\bsnm{Kundu},~\bfnm{Poorbita}\binits{P.}},
  \bauthor{\bsnm{Zhou},~\bfnm{Hang}\binits{H.}} \AND
  \bauthor{\bsnm{M{\"u}ller},~\bfnm{Hans-Georg}\binits{H.-G.}}
(\byear{2025}).
\btitle{Exceedance and force of centrality for functional data}.
\bjournal{arXiv preprint arXiv:2504.02296}.
\end{barticle}
\endbibitem

\bibitem[\protect\citeauthoryear{Kurisu et~al.}{2024}]{kurisu2024geodesic}
\begin{barticle}[author]
\bauthor{\bsnm{Kurisu},~\bfnm{Daisuke}\binits{D.}},
  \bauthor{\bsnm{Zhou},~\bfnm{Yidong}\binits{Y.}},
  \bauthor{\bsnm{Otsu},~\bfnm{Taisuke}\binits{T.}} \AND
  \bauthor{\bsnm{M{\"u}ller},~\bfnm{Hans-Georg}\binits{H.-G.}}
(\byear{2024}).
\btitle{Geodesic Causal Inference}.
\bjournal{arXiv preprint arXiv:2406.19604}.
\end{barticle}
\endbibitem

\bibitem[\protect\citeauthoryear{Lee, Mammen and Park}{2012}]{lee2012flexible}
\begin{barticle}[author]
\bauthor{\bsnm{Lee},~\bfnm{Young~K}\binits{Y.~K.}},
  \bauthor{\bsnm{Mammen},~\bfnm{Enno}\binits{E.}} \AND
  \bauthor{\bsnm{Park},~\bfnm{Byeong~U}\binits{B.~U.}}
(\byear{2012}).
\btitle{Flexible generalized varying coefficient regression models}.
\bjournal{The Annals of Statistics}
\bvolume{40}
\bpages{1906--1933}.
\end{barticle}
\endbibitem

\bibitem[\protect\citeauthoryear{Lee et~al.}{2024}]{lee2024efficient}
\begin{barticle}[author]
\bauthor{\bsnm{Lee},~\bfnm{Eun~Ryung}\binits{E.~R.}},
  \bauthor{\bsnm{Park},~\bfnm{Seyoung}\binits{S.}},
  \bauthor{\bsnm{Mammen},~\bfnm{Enno}\binits{E.}} \AND
  \bauthor{\bsnm{Park},~\bfnm{Byeong~U}\binits{B.~U.}}
(\byear{2024}).
\btitle{Efficient functional Lasso kernel smoothing for high-dimensional
  additive regression}.
\bjournal{The Annals of Statistics}
\bvolume{52}
\bpages{1741--1773}.
\end{barticle}
\endbibitem

\bibitem[\protect\citeauthoryear{Lin, Kong and Wang}{2023}]{lin2023causal}
\begin{barticle}[author]
\bauthor{\bsnm{Lin},~\bfnm{Zhenhua}\binits{Z.}},
  \bauthor{\bsnm{Kong},~\bfnm{Dehan}\binits{D.}} \AND
  \bauthor{\bsnm{Wang},~\bfnm{Linbo}\binits{L.}}
(\byear{2023}).
\btitle{Causal inference on distribution functions}.
\bjournal{Journal of the Royal Statistical Society Series B: Statistical
  Methodology}
\bvolume{85}
\bpages{378--398}.
\end{barticle}
\endbibitem

\bibitem[\protect\citeauthoryear{Lin and
  Kulasekera}{2007}]{lin2007identifiability}
\begin{barticle}[author]
\bauthor{\bsnm{Lin},~\bfnm{Wei}\binits{W.}} \AND
  \bauthor{\bsnm{Kulasekera},~\bfnm{KB}\binits{K.}}
(\byear{2007}).
\btitle{Identifiability of single-index models and additive-index models}.
\bjournal{Biometrika}
\bvolume{94}
\bpages{496--501}.
\end{barticle}
\endbibitem

\bibitem[\protect\citeauthoryear{Lin and M{\"u}ller}{2021}]{lin2021total}
\begin{barticle}[author]
\bauthor{\bsnm{Lin},~\bfnm{Zhenhua}\binits{Z.}} \AND
  \bauthor{\bsnm{M{\"u}ller},~\bfnm{Hans-Georg}\binits{H.-G.}}
(\byear{2021}).
\btitle{Total variation regularized Fr{\'e}chet regression for metric-space
  valued data}.
\bjournal{The Annals of Statistics}
\bvolume{49}
\bpages{3510--3533}.
\end{barticle}
\endbibitem

\bibitem[\protect\citeauthoryear{Lin, M{\"u}ller and
  Park}{2023}]{lin2023additive}
\begin{barticle}[author]
\bauthor{\bsnm{Lin},~\bfnm{Z}\binits{Z.}},
  \bauthor{\bsnm{M{\"u}ller},~\bfnm{H-G}\binits{H.-G.}} \AND
  \bauthor{\bsnm{Park},~\bfnm{BU}\binits{B.}}
(\byear{2023}).
\btitle{Additive models for symmetric positive-definite matrices and {L}ie
  groups}.
\bjournal{Biometrika}
\bvolume{110}
\bpages{361--379}.
\end{barticle}
\endbibitem

\bibitem[\protect\citeauthoryear{Mammen, Linton and
  Nielsen}{1999}]{mammen1999existence}
\begin{barticle}[author]
\bauthor{\bsnm{Mammen},~\bfnm{Enno}\binits{E.}},
  \bauthor{\bsnm{Linton},~\bfnm{Oliver}\binits{O.}} \AND
  \bauthor{\bsnm{Nielsen},~\bfnm{Janni}\binits{J.}}
(\byear{1999}).
\btitle{The existence and asymptotic properties of a backfitting projection
  algorithm under weak conditions}.
\bjournal{The Annals of Statistics}
\bvolume{27}
\bpages{1443--1490}.
\end{barticle}
\endbibitem

\bibitem[\protect\citeauthoryear{Matabuena and
  Petersen}{2023}]{matabuena2023distributional}
\begin{barticle}[author]
\bauthor{\bsnm{Matabuena},~\bfnm{Marcos}\binits{M.}} \AND
  \bauthor{\bsnm{Petersen},~\bfnm{Alexander}\binits{A.}}
(\byear{2023}).
\btitle{Distributional data analysis of accelerometer data from the {NHANES}
  database using nonparametric survey regression models}.
\bjournal{Journal of the Royal Statistical Society Series C: Applied
  Statistics}
\bvolume{72}
\bpages{294--313}.
\end{barticle}
\endbibitem

\bibitem[\protect\citeauthoryear{Mohs et~al.}{1997}]{mohs1997development}
\begin{barticle}[author]
\bauthor{\bsnm{Mohs},~\bfnm{Richard~C}\binits{R.~C.}},
  \bauthor{\bsnm{Knopman},~\bfnm{David}\binits{D.}},
  \bauthor{\bsnm{Petersen},~\bfnm{Ronald~C}\binits{R.~C.}},
  \bauthor{\bsnm{Ferris},~\bfnm{Steven~H}\binits{S.~H.}},
  \bauthor{\bsnm{Ernesto},~\bfnm{Chris}\binits{C.}},
  \bauthor{\bsnm{Grundman},~\bfnm{Michael}\binits{M.}},
  \bauthor{\bsnm{Sano},~\bfnm{Mary}\binits{M.}},
  \bauthor{\bsnm{Bieliauskas},~\bfnm{Linas}\binits{L.}},
  \bauthor{\bsnm{Geldmacher},~\bfnm{David}\binits{D.}},
  \bauthor{\bsnm{Clark},~\bfnm{Chris}\binits{C.}} \betal{et~al.}
(\byear{1997}).
\btitle{Development of cognitive instruments for use in clinical trials of
  antidementia drugs: additions to the {A}lzheimer's disease assessment scale
  that broaden its scope}.
\bjournal{Alzheimer Disease \& Associated Disorders}
\bvolume{11}
\bpages{13--21}.
\end{barticle}
\endbibitem

\bibitem[\protect\citeauthoryear{Moon et~al.}{2026}]{moon2026ident}
\begin{barticle}[author]
\bauthor{\bsnm{Moon},~\bfnm{Seung~Hyun}\binits{S.~H.}},
  \bauthor{\bsnm{Park},~\bfnm{Byeong~U.}\binits{B.~U.}},
  \bauthor{\bsnm{Mammen},~\bfnm{Enno}\binits{E.}} \AND
  \bauthor{\bsnm{Lee},~\bfnm{Young~Kyung}\binits{Y.~K.}}
(\byear{2026}).
\btitle{Identification and estimation of interaction effects in nonparametric
  additive regression}.
\bjournal{Biometrika}
\bvolume{113}
\bpages{Paper No. asaf074, 21}.
\end{barticle}
\endbibitem

\bibitem[\protect\citeauthoryear{Pegoraro and
  Beraha}{2022}]{pegoraro2022projected}
\begin{barticle}[author]
\bauthor{\bsnm{Pegoraro},~\bfnm{Matteo}\binits{M.}} \AND
  \bauthor{\bsnm{Beraha},~\bfnm{Mario}\binits{M.}}
(\byear{2022}).
\btitle{Projected statistical methods for distributional data on the real line
  with the {W}asserstein metric}.
\bjournal{Journal of Machine Learning Research}
\bvolume{23}
\bpages{1--59}.
\end{barticle}
\endbibitem

\bibitem[\protect\citeauthoryear{Pennec and
  Arsigny}{2013}]{pennec2013exponential}
\begin{bincollection}[author]
\bauthor{\bsnm{Pennec},~\bfnm{Xavier}\binits{X.}} \AND
  \bauthor{\bsnm{Arsigny},~\bfnm{Vincent}\binits{V.}}
(\byear{2013}).
\btitle{Exponential barycenters of the canonical {C}artan connection and
  invariant means on {L}ie groups}.
In \bbooktitle{Matrix Information Geometry}
\bpages{123--166}.
\bpublisher{Springer, Heidelberg}.
\bdoi{10.1007/978-3-642-30232-9\_7}
\bmrnumber{2964451}
\end{bincollection}
\endbibitem

\bibitem[\protect\citeauthoryear{Petersen and
  M\"uller}{2019}]{petersen2019frechet}
\begin{barticle}[author]
\bauthor{\bsnm{Petersen},~\bfnm{A.}\binits{A.}} \AND
  \bauthor{\bsnm{M\"uller},~\bfnm{H.~G.}\binits{H.~G.}}
(\byear{2019}).
\btitle{Fr\'echet regression for random objects with {E}uclidean predictors}.
\bjournal{The Annals of Statististics}
\bvolume{47}
\bpages{691-719}.
\end{barticle}
\endbibitem

\bibitem[\protect\citeauthoryear{Rawat et~al.}{2022}]{rawat2022phosphorylated}
\begin{barticle}[author]
\bauthor{\bsnm{Rawat},~\bfnm{Priyanka}\binits{P.}},
  \bauthor{\bsnm{Sehar},~\bfnm{Ujala}\binits{U.}},
  \bauthor{\bsnm{Bisht},~\bfnm{Jasbir}\binits{J.}},
  \bauthor{\bsnm{Selman},~\bfnm{Ashley}\binits{A.}},
  \bauthor{\bsnm{Culberson},~\bfnm{John}\binits{J.}} \AND
  \bauthor{\bsnm{Reddy},~\bfnm{P~Hemachandra}\binits{P.~H.}}
(\byear{2022}).
\btitle{Phosphorylated tau in Alzheimer’s disease and other tauopathies}.
\bjournal{International Journal of Molecular Sciences}
\bvolume{23}
\bpages{12841}.
\end{barticle}
\endbibitem

\bibitem[\protect\citeauthoryear{Scealy and Welsh}{2014}]{scea:14}
\begin{barticle}[author]
\bauthor{\bsnm{Scealy},~\bfnm{JL}\binits{J.}} \AND
  \bauthor{\bsnm{Welsh},~\bfnm{AH}\binits{A.}}
(\byear{2014}).
\btitle{Colours and cocktails: {C}ompositional data analysis}.
\bjournal{Australian \& New Zealand Journal of Statistics}
\bvolume{56}
\bpages{145--169}.
\end{barticle}
\endbibitem

\bibitem[\protect\citeauthoryear{Sch{\"o}tz}{2022}]{schotz2022nonparametric}
\begin{barticle}[author]
\bauthor{\bsnm{Sch{\"o}tz},~\bfnm{Christof}\binits{C.}}
(\byear{2022}).
\btitle{Nonparametric regression in nonstandard spaces}.
\bjournal{Electronic Journal of Statistics}
\bvolume{16}
\bpages{4679--4741}.
\end{barticle}
\endbibitem

\bibitem[\protect\citeauthoryear{Severn, Dryden and
  Preston}{2022}]{severn2022manifold}
\begin{barticle}[author]
\bauthor{\bsnm{Severn},~\bfnm{Katie~E}\binits{K.~E.}},
  \bauthor{\bsnm{Dryden},~\bfnm{Ian~L}\binits{I.~L.}} \AND
  \bauthor{\bsnm{Preston},~\bfnm{Simon~P}\binits{S.~P.}}
(\byear{2022}).
\btitle{Manifold valued data analysis of samples of networks, with applications
  in corpus linguistics}.
\bjournal{The Annals of Applied Statistics}
\bvolume{16}
\bpages{368--390}.
\end{barticle}
\endbibitem

\bibitem[\protect\citeauthoryear{Song and M{\"u}ller}{2026}]{song2025inference}
\begin{barticle}[author]
\bauthor{\bsnm{Song},~\bfnm{Wookyeong}\binits{W.}} \AND
  \bauthor{\bsnm{M{\"u}ller},~\bfnm{Hans-Georg}\binits{H.-G.}}
(\byear{2026}).
\btitle{Inference for dispersion and curvature of random objects}.
\bjournal{Journal of the American Statistical Association}
\bvolume{121}
\bpages{729--740}.
\end{barticle}
\endbibitem

\bibitem[\protect\citeauthoryear{Song et~al.}{2024}]{song2024multi}
\begin{barticle}[author]
\bauthor{\bsnm{Song},~\bfnm{Wookyeong}\binits{W.}},
  \bauthor{\bsnm{Oh},~\bfnm{Hee-Seok}\binits{H.-S.}},
  \bauthor{\bsnm{Cheung},~\bfnm{Ying~Kuen}\binits{Y.~K.}} \AND
  \bauthor{\bsnm{Lim},~\bfnm{Yaeji}\binits{Y.}}
(\byear{2024}).
\btitle{Multi-feature clustering of step data using multivariate functional
  principal component analysis}.
\bjournal{Statistical Papers}
\bvolume{65}
\bpages{2109--2134}.
\end{barticle}
\endbibitem

\bibitem[\protect\citeauthoryear{Stellato et~al.}{2020}]{stellato2020osqp}
\begin{barticle}[author]
\bauthor{\bsnm{Stellato},~\bfnm{Bartolomeo}\binits{B.}},
  \bauthor{\bsnm{Banjac},~\bfnm{Goran}\binits{G.}},
  \bauthor{\bsnm{Goulart},~\bfnm{Paul}\binits{P.}},
  \bauthor{\bsnm{Bemporad},~\bfnm{Alberto}\binits{A.}} \AND
  \bauthor{\bsnm{Boyd},~\bfnm{Stephen}\binits{S.}}
(\byear{2020}).
\btitle{OSQP: An operator splitting solver for quadratic programs}.
\bjournal{Mathematical Programming Computation}
\bvolume{12}
\bpages{637--672}.
\end{barticle}
\endbibitem

\bibitem[\protect\citeauthoryear{Sturm}{2003}]{sturm2003probability}
\begin{barticle}[author]
\bauthor{\bsnm{Sturm},~\bfnm{Karl-Theodor}\binits{K.-T.}}
(\byear{2003}).
\btitle{Probability measures on metric spaces of nonpositive}.
\bjournal{Heat Kernels and Analysis on Manifolds, Graphs, and Metric Spaces:
  Lecture Notes from a Quarter Program on Heat Kernels, Random Walks, and
  Analysis on Manifolds and Graphs: April 16-July 13, 2002, Emile Borel Centre
  of the Henri Poincar{\'e} Institute, Paris, France}
\bvolume{338}
\bpages{357}.
\end{barticle}
\endbibitem

\bibitem[\protect\citeauthoryear{Yang et~al.}{2025}]{yang2025variable}
\begin{barticle}[author]
\bauthor{\bsnm{Yang},~\bfnm{Haoyi}\binits{H.}},
  \bauthor{\bsnm{Bhattacharjee},~\bfnm{Satarupa}\binits{S.}},
  \bauthor{\bsnm{Xue},~\bfnm{Lingzhou}\binits{L.}} \AND
  \bauthor{\bsnm{Li},~\bfnm{Bing}\binits{B.}}
(\byear{2025}).
\btitle{Variable selection for additive global {Fr\'echet} regression}.
\bjournal{arXiv preprint arXiv:2509.13685}.
\end{barticle}
\endbibitem

\bibitem[\protect\citeauthoryear{Yu, Park and Mammen}{2008}]{yu2008smooth}
\begin{barticle}[author]
\bauthor{\bsnm{Yu},~\bfnm{KYUSANG}\binits{K.}},
  \bauthor{\bsnm{Park},~\bfnm{BYEONG~U}\binits{B.~U.}} \AND
  \bauthor{\bsnm{Mammen},~\bfnm{ENNO}\binits{E.}}
(\byear{2008}).
\btitle{Smooth backfitting in generalized additive models}.
\bjournal{The Annals of Statistics}
\bvolume{36}
\bpages{228--260}.
\end{barticle}
\endbibitem

\bibitem[\protect\citeauthoryear{Zhang, Xue and Li}{2024}]{zhang2024dimension}
\begin{barticle}[author]
\bauthor{\bsnm{Zhang},~\bfnm{Qi}\binits{Q.}},
  \bauthor{\bsnm{Xue},~\bfnm{Lingzhou}\binits{L.}} \AND
  \bauthor{\bsnm{Li},~\bfnm{Bing}\binits{B.}}
(\byear{2024}).
\btitle{Dimension reduction for Fr{\'e}chet regression}.
\bjournal{Journal of the American Statistical Association}
\bvolume{119}
\bpages{2733--2747}.
\end{barticle}
\endbibitem

\bibitem[\protect\citeauthoryear{Zhou and M{\"u}ller}{2022}]{zhou2022network}
\begin{barticle}[author]
\bauthor{\bsnm{Zhou},~\bfnm{Yidong}\binits{Y.}} \AND
  \bauthor{\bsnm{M{\"u}ller},~\bfnm{Hans-Georg}\binits{H.-G.}}
(\byear{2022}).
\btitle{Network regression with graph {L}aplacians}.
\bjournal{Journal of Machine Learning Research}
\bvolume{23}
\bpages{1--41}.
\end{barticle}
\endbibitem

\bibitem[\protect\citeauthoryear{Zhu and
  M{\"u}ller}{2023}]{zhu2023autoregressive}
\begin{barticle}[author]
\bauthor{\bsnm{Zhu},~\bfnm{Changbo}\binits{C.}} \AND
  \bauthor{\bsnm{M{\"u}ller},~\bfnm{Hans-Georg}\binits{H.-G.}}
(\byear{2023}).
\btitle{Autoregressive optimal transport models}.
\bjournal{Journal of the Royal Statistical Society Series B: Statistical
  Methodology}
\bvolume{85}
\bpages{1012--1033}.
\end{barticle}
\endbibitem

\end{thebibliography}


\end{document}